\documentclass[aps,preprint,preprintnumbers,amsmath,amssymb]{revtex4}
\usepackage{mathrsfs}
\usepackage{amssymb}
\usepackage{graphicx}
\usepackage{dcolumn}
\usepackage{bm}
\usepackage{color}
\usepackage{setspace}
\usepackage[breaklinks=true,colorlinks=true,linkcolor=blue,urlcolor=blue,citecolor=blue]{hyperref}
\UseRawInputEncoding   
\usepackage{soul}
\usepackage{ulem}
\usepackage{ragged2e}
\makeatletter

\newcommand{\Rmnum}[1]{\expandafter\@slowromancap\romannumeral #1@}

\begin{document}

\title{Rich unconventional Hall effects in a single quasi-kagome Kondo Weyl semimetal candidate Ce$_3$TiSb$_5$}

\author{Xiaobo He$^{1*}$, Ying Li$^{1,2*}$, Yongheng Ge$^{1*}$, Hai Zeng$^1$\footnote[1]{These authors contribute equally to this work.}, Shi-Jie Song$^{3}$, Jie Liu$^4$, Shuo Zou$^{1}$, Zhuo Wang$^{1}$, Yuke Li$^{5}$, Wenxin Ding$^{6}$, Jianhui Dai$^{4}$, Guang-Han Cao$^{3}$, Xiao-Xiao Zhang$^{1}$, Tianyou Zhai$^4$, Gang Xu$^{1}$\footnote[2]{Electronic address: gangxu@hust.edu.cn}, and Yongkang Luo$^{1}$\footnote[3]{Electronic address: mpzslyk@gmail.com}}

\address{$^1$Wuhan National High Magnetic Field Center and School of Physics, Huazhong University of Science and Technology, Wuhan 430074, China;}
\address{$^2$School of Science, Changzhou Institute of Technology, Changzhou 213032, China;}
\address{$^3$School of Physics, Zhejiang University, Hangzhou 310058, China;}
\address{$^4$State Key Laboratory of Materials Processing and Die \& Mould Technology, School of Materials Science and Engineering, Huazhong University of Science and Technology,Wuhan 430074, China;}
\address{$^5$School of Physics and Hangzhou Key Laboratory of Quantum Matter, Hangzhou Normal University, Hangzhou 311121, China;}
\address{$^6$School of Physics and Optoelectronics Engineering, Anhui University, Hefei 230601, China}

\date{\today}


\maketitle

\textbf{
It is generally believed that electronic correlation, geometric frustration, and topology, \textit{individually}, can facilitate the emergence of various intriguing properties that have attracted a broad audience for both fundamental research and potential applications.
Here, we report a series of unconventional Hall effects observed in a \textit{single} compound - quasi-kagome Kondo Weyl semimetal candidate Ce$_3$TiSb$_5$. In the paramagnetic phase, signature of dynamic $c$-$f$ hybridization is revealed by a reduction of anomalous Hall effect and is connected to frustration-promoted incoherent Kondo scattering. A large topological Hall effect exceeding 0.2 $\mu\Omega\cdot$cm is found at low temperatures, which should be ascribed to the non-collinear magnetic texture. In addition, a peculiar loop-shaped Hall effect with switching chirality is also seen, which is inferred to be associated with magnetic domain walls that pin history-dependent spin chirality and / or Fermi-arc surface states projected from the in-gap Weyl nodes. These exotic results place Ce$_3$TiSb$_5$ in a regime of highly-frustrated antiferromagnetic dense Kondo lattice with a nontrivial topology on an ``extended" global phase diagram, and highlight the interplay among electronic correlation, geometric frustration and topology.
}\\

\begin{figure*}[!htp]
	\vspace*{-0pt}
	\includegraphics[width=16.0cm]{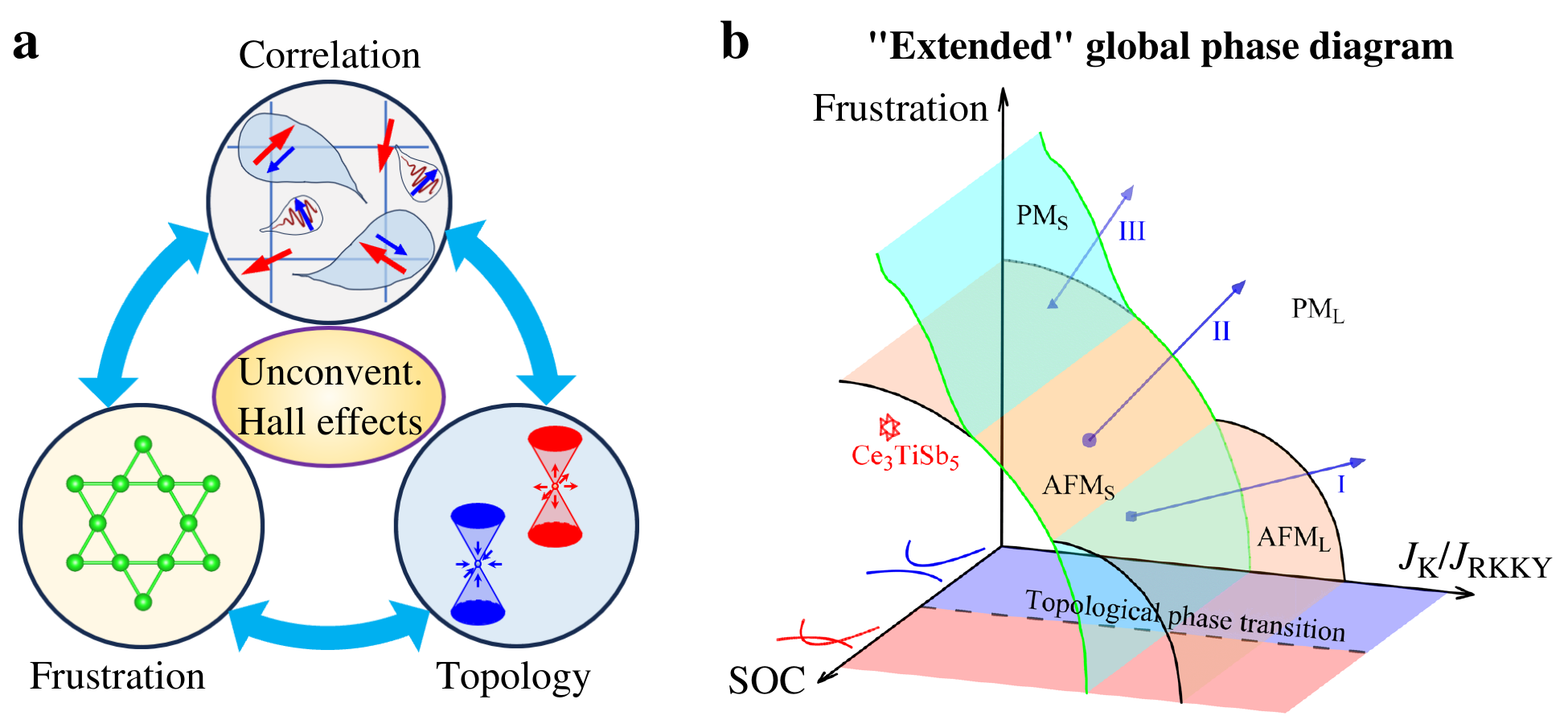}
	\label{Fig.0}
\vspace*{+10pt}
\Large
\begin{center}
    \textbf{Graphical abstract}
\end{center}
\normalsize
\end{figure*}

\clearpage

\large{\textbf{INTRODUCTION}}

Hall effect, named after Edwin Hall, is a transverse electromagnetic response that occurs when a magnetic field ($\mu_0\mathbf{H}$, referred to as $\mathbf{H}$ henceforth) is perpendicular to the current. Soon after the discovery of ordinary (or normal) Hall effect (OHE) \cite{Hall-1879}, a more overwhelming Hall signal was observed in magnetic conductors, known as anomalous Hall effect (AHE) \cite{Hall-1881}. Unlike OHE that is a consequence of Lorentz-force bending of charge-carrier trajectory induced by an \textit{external} field, the transverse velocity achieved in AHE can be of multiple origins \cite{Nagaosa-AHERMP}, from either extrinsic effect due to skew scattering \cite{Smit-1955,Smit1958} or side jump \cite{Berger-1970}, or intrinsic effect due to spin-orbit coupling (SOC) induced Berry curvature in filled bands (momentum space) \cite{KL1954,Chang-1996,Haldane-2004,Xiao-2009}. In spite of various mechanisms, a common feature of AHE is that the Hall resistivity $\rho_{yx}^{A}(H,T)$ is proportional to the magnetization $M(H,T)$. Of particular interest is that a Berry curvature can also be generated in real space when charge carriers travel across landscapes of non-coplanar or non-collinear spin textures with finite scalar spin chirality, leading to an alternative electromagnetic response - Topological Hall effect (THE) \cite{Matl-PRB1998,Ohgushi=PRB2000}. These unconventional Hall effects are attractive for both fundamental research and potential application in advanced electronic and spintronic devices.

Heavy-fermion materials are prototypical strongly correlated systems whose electronic ground states are usually determined by the competition between RKKY (Ruderman-Kittel-Kasuya-Yosida) exchange and Kondo effect \cite{Doniach}: while the RKKY exchange tends to stabilize a long-range magnetic order, the counterpart Kondo effect is to screen the local moments to form a nonmagnetic Fermi liquid. The rich physical properties of heavy-fermion compounds make them ideal material basis where singular phenomena can be found - e.g. unconventional superconductivity, quantum critical point (QCP), strange metal, etc \cite{Custers-YbRh2Si2QCP,Tsujimoto-PrV2Al20SC,Shen-CeRh6Ge4FMQCP}. Here we report a series of unconventional Hall effects in Ce$_3$TiSb$_5$, a candidate of geometrically frustrated Kondo Weyl semimetal with distorted-kagome structure. The main findings of our work are: (1) In the paramagnetic state well above the Kondo scales, the observed AHE can not be described by a regular incoherent skew scattering (or side jump, either) predicted for most Kondo lattice systems \cite{Coqblin-CeHall,Fert-SkewCe,Ramakrishnan-AHEKondo,Coleman-AHEMixed,Fert-HFAHE,Panwar-HFHall}; (2) In the magnetically ordered state, on top of the THE that is commonly seen in spin-textured systems, abnormal loop-shaped Hall effect (LHE) with switching chirality is also seen; (3) First-principles calculations suggest Ce$_3$TiSb$_5$ as a candidate of magnetic Weyl semimetal. These peculiar results suggest an extended global phase diagram for heavy-fermion systems and thus invoke further consideration about the mechanism of unconventional Hall effects in systems where topology, electronic correlation and geometric frustration meet [Fig.~1(a)]. \\

\large{\textbf{RESULTS and DISCUSSION}}

\textbf{Sample characterization}

Ce$_3$TiSb$_5$ crystallizes in the anti-Hf$_5$CuSn$_3$-type hexagonal structure with space group P6$_3$/mcm (No. 193, centrosymmetric). The crystalline structure is shown in Fig.~1(b). All the Ce atoms reside on the $z=0.25$ and 0.75 planes, and in each plane, the nearest neighbors form equilateral triangles which then build up a 2D network of distorted kagome structure, seeing also in Fig.~5(c). More details about the chemical composition and crystalline structure are provided in \textbf{Supplementary Information} (\textbf{SI}).

The sample quality of Ce$_3$TiSb$_5$ crystals was characterized by magnetic susceptibility ($\chi$) and resistivity ($\rho$), as shown in Fig.~1(c-e). Above $\sim$ 100 K, $\chi_i(T)$ ($i$ = $ab$, $c$) conforms to Curie-Weiss behavior with the fitted effective moment $\mu_{eff}=2.6(1) \mu_B$, very close to that of a free Ce$^{3+}$ ion, 2.54 $\mu_B$. This demonstrates that the 4$f$ electrons in Ce$_3$TiSb$_5$ are highly localized. According to an earlier neutron-scattering work, a commensurate antiferromagnetic (AFM) order with wave vector $\mathbf{k}_1=[0,~1/2,~1/8]$ appears below $T_{N1}\approx 5.1$ K; further cooled down, a second related AFM order ($T_{N2}\sim 3$ K) with $\mathbf{k}_2=[0,~0,~1/8]$ slowly develops with time in the presence of the $\mathbf{k}_1$ phase, and the coexistence of $\mathbf{k}_1$ and $\mathbf{k}_2$ phases persists even for well below $T_{N1}$ \cite{Ritter-JPCM2021}. The two proposed magnetic structures are displayed in Fig.~1(b). In our magnetic susceptibility measurements, the AFM transition at $T_{N1}$ is clearly seen in $\chi(T)$, whereas the feature at $T_{N2}$ is tiny and can be visible only in derivative [inset to Fig.~1(c)]. The anisotropic $\chi(T)$ profiles also manifest $\mathbf{ab}$ plane as the easy plane, consistent with neutron results and can be well understood by the crystal electric field (CEF) effect \cite{Matin-JPCM2017}. Field dependent magnetization for $\mathbf{H}\parallel[1000]$ at 2 K displays multiple metamagnetic transitions, manifested by the discontinuous jumps as shown in Fig.~1(d). When field is along $[10\bar{1}0]$, $M_{1/3}$ and $M_{2/3}$ plateaus are observable, reminiscent of the kagome spin ice state (2-in-1-out or 1-in-2-out) in HoAgGe \cite{Zhao-HoAgGe}. Such metamagnetic transitions likely are consequences of the competition among magnetic exchange, Zeeman energy and geometric frustration. The magnetization for $\mathbf{H}\parallel[0001]$ is much smaller, and is essentially linear with field, reaffirming $\mathbf{c}$ as hard axis. It should be pointed out that since the local centrosymmetry is broken at Ce ($m2m$) sites in Ce$_3$TiSb$_5$, Dzyaloshinskii-Moriya interaction that arises from non-centrosymmetric SOC would also favor $\mathbf{ab}$ as magnetic easy plane.

Temperature dependence of resistivities are shown in Fig.~1(e). For both $\mathbf{I}\parallel[1000]$ and [0001], the $\rho(T)$ curves exhibit clear humps near 100 K, characteristic of scatterings by excitation between CEF levels. A local minimum can be seen at $T_K^{on} \sim 30$ K, the signature for the onset of incoherent Kondo scattering; and this is followed by a pronounced maximum in $\rho$ near $T_{coh}\sim 10$ K below which coherent Kondo liquid comes into being. Previous specific heat measurements revealed an enhanced Sommerfeld coefficient $\sim$600 mJ/mol$\cdot$K$^2$ (compared to 9.0 mJ/mol$\cdot$K$^2$ for La$_3$TiSb$_5$), confirming substantial electronic correlation effect. The single-ion Kondo scale was estimated $T_K\sim 8$ K according to the magnetic entropy \cite{Matin-JPCM2017}. All together, Ce$_3$TiSb$_5$ appears as a frustrated AFM Kondo lattice with moderate electronic correlation. Noteworthy that a tiny trace of superconductivity is visible in $\rho(T)$ around 3.5 K, which should be attributed to a small amount of Sn-flux impurity \cite{Matin-JPCM2017}.\\

\textbf{AHE in the paramagnetic phase}

Figure~2(a) displays the field dependence of Hall resistivity $\rho_{yx}$ at various temperatures. In these Hall effect measurements, the electrical current flows within the $\mathbf{ab}$ plane, and the magnetic field was applied along [0001], the magnetic hard axis [cf. the inset to Fig.~2(a)]. For all the temperatures measured between 2 K and 300 K, the Hall signal is negative. In the paramagnetic phase ($T\geq 7$ K), $\rho_{yx}$ is essentially linear with $H$. (Linear $\rho_{yx}(H)$ was also observed in La$_3$TiSb$_5$; all these indicate that multiband effect is not significant in the ordinary Hall effect for (La,Ce)$_3$TiSb$_5$.) At first glance, Hall coefficients can be extracted straightforwardly from the slope of $\rho_{yx}(H)$, as shown in Fig.~2(b). However, we notice that the obtained $\rho_{yx}/H$ exhibits a significant temperature dependence, rather unusual for a conventional metal. In particular, $\rho_{yx}/H$ appears inversely proportional to $T$ above 30 K (i.e., Curie-Weiss-like), below which the magnitude of $\rho_{yx}/H$ falls back. The temperature where $|\rho_{yx}/H|$ maximizes coincides with $T_K^{on}$ observed in $\rho(T)$ [seeing Fig.~1(c)]. This implies that the asymmetric scattering by magnetic ions is dominant here even in the paramagnetic phase. A conventional expression of Hall resistivity in magnetic conductors consists of two items, $\rho_{yx}=R_H H + R_A M(H)$, where $R_H$ is the normal Hall coefficient, and the second term is due to AHE. (Note that Ce$_3$TiSb$_5$ is a multi-band system, we here avoid relating $R_H$ directly to the carrier density as in single-band cases.) Earlier studies on AHE have revealed that the form of $R_A$ is typically in the form of $A \rho^{\beta}$, where $A$ is a prefactor, and the exponent $\beta$ depends on specific mechanisms \cite{Nagaosa-AHERMP}: $\beta\approx 1$ for skew scattering, and $\beta\approx 2$ for intrinsic and side-jump scattering. We, therefore, describe $\rho_{yx}/H$ as follows,
\begin{equation}
    \rho_{yx}/H=R_H + A \rho^{\beta} \chi_c,
    \label{Eq.1}
\end{equation}
and this is better expressed in a plot of $\rho_{yx}/H$ vs. $\rho^{\beta}\chi_c$ with $T$ as an implicit parameter, as shown in the inset to Fig.~2(b). We find that only $\beta\approx 0$ fits the data satisfactorily well for $T\geq T_K^{on}$ [dash line in the inset to Fig.~2(b)], while both $\beta=1$ and 2 fail in the fitting (Supplementary Figure 3 in \textbf{SI}). The extrapolation of this fitting to the $\chi_c\rightarrow0$ limit results in the normal Hall coefficient of Ce$_3$TiSb$_5$, $R_H=+0.6(2)\times10^{-10}$ m$^3$/C, meaning that the system actually is hole-like. It should be pointed out that the Hall coefficients of the non-magnetic reference La$_3$TiSb$_5$ are $+0.42\times10^{-10}$ m$^3$/C at 2 K and $+0.60\times10^{-10}$ m$^3$/C at 300 K, in close agreement with that of Ce$_3$TiSb$_5$ in both sign and magnitude, implying the validity of our analysis. The reason for the negative Hall response in Ce$_3$TiSb$_5$ is because the small positive OHE is buried by a large negative AHE. It is worthwhile to mention that opposite signs of OHE and AHE in Ce-contained compounds seem rather common, e.g. our previous work on CeNi$_{2-\delta}$As$_{2}$ revealed a positive AHE albeit the dominant carriers are electrons \cite{LuoY-CeNi2As2Pre}.

Anomalous behavior of Hall effect has been extensively studied in many heavy-fermion Kondo lattices, like CeAl$_3$ \cite{Brandt-CeAl3Hall}, CeCu$_6$ \cite{Onuki-CeCu6}, CeRu$_2$Si$_2$ \cite{Leroux-HallHF}, Ce(Co,Rh,Ir)In$_5$ \cite{Hundley-Ce115AHE}, etc \cite{Lapierre-HFHall, LuoY-CeNi2As2Pre}. Unlike conventional magnetic conductors, in heavy-fermion materials, the hybridization between localized $f$-electrons and conduction ($c$) electrons leads to the Kondo effect \cite{Hewson-Kondo} the consequence of which is to screen the local moments. Such a screening may start to set in as a local, incoherent form below the onset Kondo temperature $T_K^{on}$ [Fig.~2(c)]. Further cooled down to $T<T_{coh}$, the Kondo effect becomes coherent by showing a renormalized Kondo resonance near the Fermi level, and the ground state of the system turns into a Fermi liquid with large quasiparticle effective mass. Due to the coherent Kondo screening, the asymmetric scattering at the magnetic ions is expected to be reduced, and correspondingly, the strength of AHE will be affected. Previous experimental works revealed a universal temperature dependence in $|\rho_{yx}/H|$: it increases substantially with decreasing temperature, passing through a maximum near $T_{coh}$, and then decreases rapidly below $T_{coh}$ \cite{Brandt-CeAl3Hall,Leroux-HallHF,Onuki-CeCu6}. A variety of theoretical models based on modified skew scattering have been proposed for the AHE in heavy-fermion Kondo lattices \cite{Ramakrishnan-AHEKondo,Coleman-AHEMixed,Fert-HFAHE,Panwar-HFHall,Kohno-JMMM1990,Yamada-PTP1993,Kontani-AHEHF}, seeing Table 1.

On the whole, our AHE results in Ce$_3$TiSb$_5$ are at odds with all the existing theories for heavy-fermion systems. The obtained $\rho_{yx}/H\propto\chi\rho^0$ shown in Fig.~2(b) resembles Kontani \textit{et al}'s prediction at high temperature on the basis of the periodic Anderson Hamiltonian with degenerate orbitals \cite{Kohno-JMMM1990,Yamada-PTP1993,Kontani-AHEHF}, however, we notice that this relation is ended up near $T_K^{on}$, a temperature that is about 3 times of $T_{coh}$ [Tab.~1]. In other words, the suppression of asymmetric magnetic scattering appears prior to the establishment of collective Kondo coherence, which is reminiscent of the hybridization fluctuations (or dynamic Kondo hybridization) in CeCoIn$_5$ observed by angle-resolved photoemission spectroscopy \cite{Chen-CeCoIn5ARPES} and ultrafast optical pump-probe spectroscopy \cite{Liu-CeCoIn5HybriFluc} measurements. \textit{Our results provide a remarkable sign for this region by anomalous Hall effect which has not been reported before.} It is likely that geometric frustration adds to the Kondo effect (an inherent frustration to magnetic order), and promotes the incoherent Kondo scattering, reducing local magnetization and thus suppressing the AHE even without Kondo coherence. A schematic illustration is provided in Fig.~2(c-d). Indeed, geometric frustration has been found to play a crucial role in the physical properties and quantum phase diagrams of heavy-fermion systems. For instance, in CePdAl that is also quasi-kagome structured (Supplementary Figure 9), the involvement of geometric frustration gives rise to a peculiar quantum-critical phase \cite{Zhao-CePdAlQCP} that was theoretically predicted to emerge at the high-frustration regime in the ``Global phase diagram" of heavy-fermion systems \cite{Si-Global,Coleman-Frustration}, seeing also the path III in Fig.~6. However, we must admit that such a scenario of frustration-promoted dynamic Kondo hybridization still needs to be testified in more heavy-fermion materials, and also that an appropriate description about the influence of geometric frustration on the AHE in heavy-fermion systems is still lacking. Our work strongly suggests this necessity in the future.\\

\textbf{Unconventional Hall effects in the AFM phase}

The low-temperature part of Hall effect is displayed in Fig.~3(a). When $T\leq7$ K, $\rho_{yx}$ becomes superlinear with $H$. Because $M(H)$ is sublinear [cf. Fig.~3(b)], the superlinear $\rho_{yx}(H)$ can not be simply understood by a standard AHE, but instead, the high-field part of $\rho_{yx}(H)$ can be well fit to a $H^2$ law; as $T$ decreases, this quadratic regime extends to lower fields; seeing the dashed lines in Fig.~3(d). Such a temperature and field dependence mimics the situation in CeIrIn$_5$ \cite{Nair-CeIrIn52008} and
is anticipated as the high-field limit for compensated metals, viz $\omega_c\tau \gg 1$ where $\omega_c\equiv eH/m^*$ is the cyclotron frequency, and $\tau$ is the scattering time \cite{Hurd-HallEffect}. Another important feature of $\rho_{yx}(H)$ is that a loop-shaped component (LHE) appears at low temperature especially for below $T_{N2}$, i.e. the $\rho_{yx}(H)$ recorded when field swept from $+14$ T to $-14$ T ($\rho_{yx}^{+-}$) does not overlap with that from $-14$ T to $+14$ T ($\rho_{yx}^{-+}$). What attracts us even more is that at 2 K, the base temperature of our measurements, we see the loop twists for multiple times, or in other words, the emergent sub-loops are of different chiralities. Such a peculiar LHE has not been observed in other compounds, so far as we know. We tentatively express the total Hall resistivity as follows,
\begin{equation}
    \rho_{yx}(H)=\rho_{yx}^O+\rho_{yx}^A+\rho_{yx}^T+\rho_{yx}^L,
\end{equation}
where $\rho_{yx}^O$, $\rho_{yx}^A$, $\rho_{yx}^T$ and $\rho_{yx}^L$ stand for ordinary / anomalous / topological / loop-shaped Hall resistivities, respectively. First, the LHE component can be easily separated out by $\rho_{yx}^L(H)=\rho_{yx}^{-+}-\rho_{yx}^{+-}$, and the resultant are presented in Fig.~3(c). The sum of the rest components ($\rho_{yx}-\rho_{yx}^L$) is shown in Fig.~3(d), together with the $H^2$ fitting (dashed line) mentioned here above. One can clearly see anomalous bumps on top of the smooth $H^2$ background in $\rho_{yx}-\rho_{yx}^L$. Since such bumps are absent in $M(H)$ [Fig.~3(b)], they cannot be attributable to a standard AHE \cite{Nagaosa-AHERMP}, but should be categorized as THE. We extract the THE component $\rho_{yx}^T$ by subtracting the $H^2$ background from $\rho_{yx}-\rho_{yx}^L$, and the obtained are shown in Fig.~3(e). Similar results were obtained in other samples [Fig.~3(f) and Supplementary Figure 5]. The maximum of $\rho_{yx}^T$ exceeds 0.2 $\mu\Omega\cdot$cm at 2 K, which is comparable with that of MnGe \cite{Kanazawa-MnGeTHE}, and is two orders of magnitude larger than in MnSi \cite{Neubauer-MnSiTHE,Schulz-MnSiTHE}. It is unlikely that such a large THE arises from some extrinsic mechanism such as inhomogeneous current paths, seeing Supplementary Figure 5. We should note that the components $\rho_{yx}^O$ and $\rho_{yx}^A$ in this region are hardly dissolvable; a multiband effect may be also present to cause the nonlinear $\rho_{yx}$ with respect to field; however, whether $\rho_{yx}^O$ and $\rho_{yx}^A$ can be separated, or if multiband effect is indeed active, has little impact on the extracted THE component.

To discuss about the origin of these peculiar Hall effects seen inside the AFM phase. Apparently, both $\rho_{yx}^T$ and $\rho_{yx}^L$ are of magnetic origin in that these signals disappear right above $T_{N1}$. Meanwhile, in any case one can not scale $\rho_{yx}(H)$ with $M(H)$, suggesting the existence of components other than the standard AHE. Therefore, it is reasonable to attribute the aforementioned $\rho_{yx}^T$ and $\rho_{yx}^L$ to the so-called THE. The past decades have witnessed a boost of THE observed in serial materials with non-coplanar or non-collinear spin texture, such as Mn(Si,Ge) \cite{Lee-MnSiTHE,Neubauer-MnSiTHE,Schulz-MnSiTHE,Kanazawa-MnGeTHE}, GdPtBi \cite{Suzuki-GdPtBiTHE}, MnBi$_4$Te$_7$ \cite{Roychowdhury-MnBi4Te4THE}, Gd$_2$PdSi$_3$ \cite{Kurumaji-Gd2PdSi3THE}, YMn$_6$Sn$_6$ \cite{Ghimire-YMn6Sn6THE}, and EuGa$_2$Al$_2$ \cite{Moya-EuGa2Al2THE}. This scenario is likely also applicable to Ce$_3$TiSb$_5$, considering the magnetic structures proposed by neutron diffraction \cite{Ritter-JPCM2021}. For both $\mathbf{k}_1$ and $\mathbf{k}_2$ phases of Ce$_3$TiSb$_5$, the moments of Ce are aligned within the $\mathbf{ab}$ plane, and the angles spanned by the nearest neighbors are either 120 $^\circ$ or 60 $^\circ$ [Fig.~1(b)]. Under field $\mathbf{H} \parallel$ [0001], these moments are out-of-plane canted and become both non-collinear and non-coplanar; a finite scalar spin chirality $\chi_{ijk}=\mathbf{S}_i\cdot(\mathbf{S}_j\times\mathbf{S}_k)$ hence is accomplished. This makes the observation of THE below $T_{N1}$ rather likely. We are impressed that the observed THE here appears across a broad field range, likely due to the slow polarization process when field is applied along the hard axis of Ce$_3$TiSb$_5$.

To the best of our knowledge, the loop-shaped Hall effect - which in our opinion is a special kind of THE - has only been reported in a handful materials including Nd$_{2}$Ir$_2$O$_7$ \cite{Disseler-Nd2Ir2O7LHE,Ueda-Nd2Ir2O7LHE} and CeAl(Si,Ge) \cite{Yang-CeAlSiLHE,Piva-CeAlSiPressure,He-CeAlGeLTHE}. A major commonality of them is that they are all candidates of magnetic Weyl semimetal with domain wall (DW) in bulk. In the pyrochlore-type AFM insulator Nd$_{2}$Ir$_2$O$_7$, the DWs at the boundary between all-in-all-out and all-out-all-in domains pin in-gap Weyl fermions and the projected Fermi-arc surface states which give rise to an emergent metallic interface \cite{Yamaji-MetallicDW} and a loop-shaped electromagnetic response \cite{Ma-Nd2Ir2O7}. Our previous work on CeAlGe also suggested an alternative scenario by taking into account the spin chirality in the magnetic DWs \cite{He-CeAlGeLTHE}. Since DWs are a consequence of meta-stable phases coexisting in the bulk, the magnetization in the DW is usually history dependent. In this context, the loop-shaped Hall response arising from DWs can be understood as follows. Previous neutron experiment revealed two related magnetic orders $\mathbf{k}_1$ and $\mathbf{k}_2$ in Ce$_3$TiSb$_5$ below $T_{N2}$. More importantly, the transition between $\mathbf{k}_1$ and $\mathbf{k}_2$ seems 1st order with significant hysteresis and phase coexistence \cite{Ritter-JPCM2021}. DWs are expected to form between these different magnetic phases, which potentially leads to the history dependent real-space Berry curvature and hence the loop-shaped Hall response. A schematic diagram of this scenario is present in Fig.~4(a). Since these DWs are interfaces, conceivably they are hardly detectable by bulk measurements like magnetic susceptibility. Sample dependence is another feature of DWs, because they are usually pinned by impurities or defects in the crystal. This is indeed true for the LHE in other samples, as shown in Fig.~3(f) and Supplementary Figure 5. The scenario can be also supported by the Magnetic-Force Microscopy (MFM) measurements that image the presence of magnetic domains at low temperature for field up to 9 T, seeing Supplementary Figure 6. However, we must admit that to accurately account for the multiple twists observed in $\rho_{yx}^L(H)$ at 2 K [Fig.~3(a,f)], more local, quantitative probes to this interface spin chirality will be needed, which is far beyond our ability at present. Imaging the DW structures by microscopic techniques like vector magneto-optical Kerr effect (VMOKE) and scanning superconducting quantum interference device (SQUID) may be informative \cite{Xu-CeAlSiDW,Sun-CeAlSiDW}.\\

\textbf{DFT calculations}

It is interesting to note that the observed unconventional Hall effects in Ce$_3$TiSb$_5$ can be also mediated by nontrivial band topology. This is partly motivated by the topological property in perfect kagome lattices (e.g., see a recent review \cite{Yin-kagomeReview2022} and the references therein), and can be supported by the DFT calculations. We start from La$_3$TiSb$_5$ that can be deemed as the non-magnetic reference of Ce$_3$TiSb$_5$. The derived Fermi surface and band structure of La$_3$TiSb$_5$ are in Fig.~5(a-b). In the presence of both time reversal ($\mathcal{T}$) and inversion ($\mathcal{P}$) symmetries, all bands are doubly degenerate. Three doubly degenerated bands are found to pass through the Fermi energy ($E_F$), labeled as 111 (blue), 113 (red) and 115 (red). Band 111 has multiple crossing points with $E_F$, while bands 113 and 115 are dispersive only along $\Gamma$-$\text{Z}$. Correspondingly, the Fermi surfaces show two different topologies, a 3D-like open pocket enclosing the $\Gamma$ point, and two flat sheets originating from the quasi-1D 113 and 115 bands. Of prime interest is that the bands 113 and 115 cross with each other linearly and give rise to a tilted Dirac node that is about 47 meV above $E_F$. La$_3$TiSb$_5$, therefore, is likely a type-\Rmnum{2} Dirac semimetal. It is worthwhile to mention here that La$_3$MgBi$_5$, isostructural to La$_3$TiSb$_5$, was also suggested as a topological Dirac semimetal by quantum oscillations and DFT calculations \cite{HanX-La3MgBi5}.

Turning now to the band structure of Ce$_3$TiSb$_5$. It is well known that by breaking either $\mathcal{T}$ or $\mathcal{P}$ symmetry, a Dirac point splits into two Weyl points with opposite chiralities \cite{Weyl1929,Weng-TmPn}. Regarding to the situation in Ce$_3$TiSb$_5$, it is extremely difficult to calculate its band structure by taking into account the $\mathbf{k}_1$ or $\mathbf{k}_2$ type magnetic structure. Instead, a simpler magnetic unit cell without $\mathbf{c}$-axis modulation was considered in the DFT calculations. Since all the Ce atoms reside in two individual planes that are separated by nonmagnetic Ti-Sb slabs, such a treatment maintains the time-reversal symmetry breaking without qualitatively altering the band topology, seeing EuCuAs \cite{Roychowdhury-EuCuAsJACS2023} and MnBi$_2$Te$_4$ \cite{Tan-MnBi2Te4PRL2020} for instances. To make the results closer to the $\mathbf{k}_2$ ground state \cite{Ritter-JPCM2021}, we imposed two prerequisites: (i) all Ce moments lay inside the $\mathbf{ab}$ plane, and (ii) a ferromagnetic interchain coupling exists between the two Ce sublattices. Firstly, a moderate Coulomb $U=2$ eV and Hund¡¯s coupling $J=0.5$ eV were employed. The calculation manifests many magnetic phases with close energy scales (Supplementary Table 3 and Supplementary Figure 7 in \textbf{SI}), in line with the magnetic frustration exerted by the quasi-kagome lattice. A brief summary for the calculations with different magnetic structures is listed in Table 2. The configuration with lowest energy (termed as AFM$^{\text{F}}_{60/120}$) is shown in Fig.~5(c), which essentially captures all the features of $\mathbf{k}_2$ only except for the $\mathbf{c}$-axis modulation. In this magnetic structure, there are two moments pointing inside and one moment outside (viz. 2-in-1-out) in one triangle and 1-in-2-out in the adjacent triangle, and the angle between nearest moments is either 60$^{\circ}$ or 120$^{\circ}$; for each Ce6 polyhedron, a net magnetization is found to be along $[10\bar{1}0]$, cf. the orange arrow in Fig.~5(c), breaking the $\mathcal{T}$ symmetry. The band structure of AFM$^{\text{F}}_{60/120}$ is shown in Fig.~5(d). $c$-$f$ hybridization is obviously seen, and narrow bands with substantial Ce-$4f$ character are found just below $E_F$, qualitatively consistent with the moderate electronic correlation revealed by specific heat. Due to the $\mathcal{T}$ symmetry breaking, each doubly degenerated band is split. Two Weyl nodes (denoted by WP1 and WP2) appear on $\Gamma$-$\text{Z}$ line at the crossing points of the bands with different irreducible representations (LD$_3$ and LD$_4$, respectively). To test the robustness of this topological feature to electronic correlation, we considered the situations with different $U$, and the results are displayed in Supplementary Figure 8 in \textbf{SI}. As for smaller $U=1$ eV, the narrow bands of Ce-4$f$ cross $E_F$, implying partial delocalization of $4f$ electrons and that the system probably enters the intermediate-valence regime. For large $U=6$ eV, the Ce-$4f$ electrons are fully localized, and the band structure appears similar to that of La$_3$TiSb$_5$, whereas the $\mathcal{T}$ breaking induces a tiny splitting of each band. It is worthwhile to emphasize that in spite of the different energy levels of 4$f$ bands, the Weyl points survive on $\Gamma$-$\text{Z}$. Therefore, the ground state of Ce$_3$TiSb$_5$ is likely to be a type-\Rmnum{2} Weyl semimetal.

For comparison, we also attempted to calculate the band structure of Ce$_3$TiSb$_5$ with $\mathbf{k}_1=[0,~1/2,~1/8]$ magnetic state. The magnetic unit cell in this case is doubly expanded along $\mathbf{b}$. Here, again, we ignored the $\mathbf{c}$-axis modulation, and the simplified magnetic structure (termed as AFM-Zigzag$^\text{F}$) is shown in Fig.~5(e). The calculated band structure with $U=2.0$ eV and $J=0.5$ eV is displayed in Fig.~5(f). In this magnetic structure, the Zigzag-arranged Ce moments cause a net moment $M_x$ in a Ce$_6$ polyhedron, whereas the adjacent Ce$_6$ polyhedra along $\mathbf{b}$ follow a sequence of $M_x$, $M_x$, $-M_{x}$, $-M_{x}$ in one expanded unit cell; on the whole, the $\mathcal{PT}$ symmetry preserves. However, the little group (2$^\prime$/m) of $\Gamma$-Z cannot protect the Dirac node, which gives the gapped bands along $\Gamma$-Z, comparing to the Weyl nodal bands in $\mathbf{k}_2$ phase.
All together, the $\mathbf{k}_1$ and $\mathbf{k}_2$ phases are potentially of different topological characters. In this context, one may expect that the domain wall between the $\mathbf{k}_1$ and $\mathbf{k}_2$ phases can pin chiral Fermi arc that connects two Weyl points with opposite chiralities [Fig.~4(b)]. Such a metallic interface may give a second explanation for the observed loop-shaped Hall effect.\\

\textbf{Additional remarks}

On the example of Ce$_3$TiSb$_5$, we clearly see the interplays among electronic correlation, topology and geometric frustration. Compared with earlier Kondo Weyl semimetal candidates like Ce$_3$Bi$_4$Pd$_3$ \cite{Dzsaber-Ce3Bi4Pd3}, YbPtBi \cite{Guo-YbPtBiKWSM} etc, \textit{the involvement of geometric frustration sheds new lights to correlated topological materials}. On the one hand, geometric frustration can promote the inherent frustration exerted by Kondo coupling, and further weakens the asymmetric skew scattering prior to the establishment of Kondo coherence; on the other hand, it also leads to finite scalar spin chirality and domain walls between quasi-degenerate magnetic orders. All these are reflected in our Hall effect studies.

The frustration-mediated textured AFM orders in Ce$_3$TiSb$_5$ remind us of another possible mechanism for the AHE above $T_{N1}$. According to H. Ishizuka \textit{et al} \cite{HIshizuka-SSCSkew}, an alternative type of skew-scattering via thermally excited spin-clusters with scalar spin chirality can cause a giant AHE whose Hall angle ($\theta_H^A$) could be as large as $0.1\pi$, and this has been exemplified in the chiral magnet MnGe \cite{YFujishiro-MnGeSSCSkew}. To discuss about this possibility, we show $\tan\theta_H^A$ as a function of $T$ in Supplementary Figure 4. The Hall angle here is less than 1\%, implying that the scenario of spin-chirality skew scattering originally proposed for weakly-correlated metals is likely non-applicable here, but counter-indicating the important role played by Kondo hybridization \cite{SiddiqueeH-USbTeAHE}.

Finally, all these unconventional Hall effects observed in Ce$_3$TiSb$_5$ motivate us to discuss further about the emergent physics on the standpoint of heavy-fermion system. This might be sparked by proposing an ``extended" global phase diagram for heavy-fermion systems, as shown in Fig.~6. The classic global phase diagram \cite{Si-Global} was proposed to describe the classification of quantum phase transitions in heavy-fermion materials with various frustration parameter ($G$) and ratio between Kondo coupling and RKKY interaction ($J_K/J_{RKKY}$). The green line represents Kondo-breakdown transition across which the highly localized $f$ electrons become delocalized, and this is accompanied with an abrupt change in Fermi surface. By convention, we here use the subscripts ``S" or ``L" to depict the magnetic states with small or large Fermi surface. Previous studies suggested that Kondo-breakdown type QCP typically occurs in systems with relatively high $G$ (path-II, AFM$_\text{S}$-PM$_\text{L}$), while sequential Kondo-breakdown transition and spin-density-wave (SDW) type QCP were found more often in systems with lower $G$ (path-I, AFM$_\text{S}$-AFM$_\text{L}$-PM$_\text{L}$). Some representative examples include Ce(Cu,Au)$_6$\cite{Lohneysen-CeCu6_AuQCP}, YbRh$_2$Si$_2$\cite{Custers-YbRh2Si2QCP}, Ce$_mM_n$In$_{3m+2n}$\cite{Shishido-CeRhIn5dHvA,LuoY-CeRhIn5_IrQCP,Sebastian-CeIn3}, CeNiAsO\cite{LuoY-CeNiAsOQCP}, etc. In a high-frustration limit, a less common quantum phase transition may take place (path-III, AFM$_\text{S}$-PM$_\text{S}$-PM$_\text{L}$), where the magnetic order quenches prior to Kondo-breakdown transition, and the Fermi surface reconstructs within the paramagnetic state. Recently, this rare case has been exemplified by CePdAl tuned by both pressure and field \cite{Zhao-CePdAlQCP}. In CePdAl, the high frustration is fulfilled by the quasi-kagome structure, as the case in our Ce$_3$TiSb$_5$. The topological features observed in Ce$_3$TiSb$_5$ employ topology - a new degree of freedom that can be switched by SOC - to the classic global phase diagram. Our results place Ce$_3$TiSb$_5$ at the highly-frustrated AFM$_\text{S}$ regime with non-trivial topology. With this in mind and also taking into account the rich magnetic and topological phases proposed by DFT, natural questions can be further put forward concerning how the ground state evolves when tuned by pressure, chemical doping, or strong magnetic field, and whether a topological phase transition can entangle with a magnetic quantum phase transition. All these await to be clarified in the future.\\

\large{\textbf{CONCLUSION}}

In all, a series of unconventional Hall effects are observed in a single quasi-kagome Kondo Weyl semimetal candidate Ce$_3$TiSb$_5$, suggestive of the intriguing interplay among electronic correlation, topology, and geometric frustration. Our work, therefore, provides a rare paradigm of correlated topological material residing in the regime of highly-frustrated AFM$_{\text{S}}$, and enriches the classic global phase diagram for heavy-fermion systems.\\

\textbf{METHODS}

\textbf{Single crystals preparation and measurements}

Single crystals of Ce$_{3}$TiSb$_{5}$ were grown by a Sn-flux method \cite{Matin-JPCM2017}. Starting materials Ce : Ti : Sb : Sn = 3 : 1 : 5 : 20 in molar ratio were weighted and placed in an alumina crucible. The  mixture was heated up to 1100 $^{\circ}$C in 5 h, held for 24 h and then slowly cooled at a rate of 1 $^{\circ}$C/h to 650 $^{\circ}$C at which temperature the Sn flux was effectively removed by centrifugation. The obtained crystals are of hexagonal prism shape with typical dimensions $1 \times 1\times 5$ mm$^{3}$. The chemical composition, the quality of crystallization, and the orientation of the measured sample were verified by Energy-dispersive X-ray spectroscopy (EDS), Wavelength-dispersive X-ray (WDX), single-crystal X-ray diffraction and Laue X-ray diffraction; the results are displayed in Supplementary Figures 1-2 and Supplementary Tables 1-2 of \textbf{SI}. Magnetization measurements were carried out in a commercial physical-property measurement system (PPMS-16 T, Quantum Design). For the transport measurements, both in-plane longitudinal resistivity ($\rho_{xx}$) and Hall resistivity ($\rho_{yx}$) were measured by a lock-in amplifier (SR865A, Standford Research) equipped with SR554A pre-amplifier. The charge current $\mathbf{I}$ was applied parallel with [1000]. For all these measurements, the magnetic field was applied along [0001]. Four samples were measured for Hall effect, labeled as S1-4, respectively. Magnetic-Force Microscopy (MFM) images were obtained using an Attocube attoDRY2100 closed-cycle cryogenic microscope with a base temperature of 2 K and a magnetic field up to 9 T.

\textbf{First-principles calculations}

First-principles calculations were performed within the density-functional-theory (DFT) formalism as implemented in VASP with the projector augmented wave method. The energy cutoff for wave-function expansion was set as 450~eV, and the $\mathbf{k}$-point sampling was 9 $\times$ 9 $\times$ 12. The generalized gradient approximation (GGA) of Perdew-Burke-Ernzerhof (PBE) for exchange-correlation potential was used in all calculations. The maximally localized Wannier functions were constructed using the WANNIER90 package based on the First-principles calculations. To account for the correlation effects of the Ce-$4f$ electrons, we have adopted GGA + $U$ method with Hubbard $U = 2$ eV and Hund's coupling $J = 0.5$ eV, which matches the moderate electronic correlation revealed by specific heat. We have also tested different $U$ values, and the results of $U = 1$ and 6 eV are shown in \textbf{SI}. SOC was implemented in all calculations. \\

\textbf{Data availability}

The data supporting the findings of this study are available from the corresponding authors upon reasonable request via email to Y. Luo. \\

\emph{}\\

\textbf{ACKNOWLEDGEMENTS}

We are grateful to Qimiao Si, Marc Janoschek, Shizeng Lin, Wolfgang Josef Simeth, and Shangshun Zhang for insightful discussions. This work is supported by National Key R\&D Program of China (2023YFA1609600, 2022YFA1602600, 2024YFA1611200, 2023YFA1406101), National Natural Science Foundation of China (U23A20580, 12274154, 12404182), Beijing National Laboratory for Condensed Matter Physics (2024BNLCMPKF004), Hangzhou Joint Fund of the Zhejiang Provincial Natural Science Foundation of China (LHZSZ24A040001), and Guangdong Basic and Applied Basic Research Foundation (2022B1515120020). The computations were completed in the HPC Platform of Huazhong University of Science and Technology.
\\

\textbf{AUTHOR CONTRIBUTIONS}

Y.Luo conceived and designed the experiments. X.H. and H.Z. grew the crystals, characterized the samples, and performed most of the measurements. Ying L. and Y.G. carried out the first-principles calculations under the supervision of G.X. S.-J.S., S.Z., Z.W. made some of the other experimental measurements. J.L. and T.Z. made the Magnetic-Force Microscopy measurements.
Yuke L., W.D., J.D., G.-H.C., X.-X.Z. provided constructive discussions. X.H., Ying L., Y.G., H.Z., G.X. and Y.Luo. discussed the data, interpreted the results, and wrote the paper with input from all the authors. \\

\textbf{COMPETING INTERESTS}

The authors declare no competing interests.\\

\textbf{ADDITIONAL INFORMATION}

\textbf{Supplementary information} The online version contains supplementary material available at https://doi.org/10.1038/s43246-025-00927-8. \\

\textbf{Correspondence} and requests for materials should be addressed to Gang Xu or Yongkang Luo. \\

\textbf{Peer review information} Communications Materials thanks the anonymous reviewers for their contribution to the peer review of this work. \\

\textbf{Reprints and permission information} is available at \\
http:$\backslash\backslash$www.nature.com/reprints \\

\textbf{Publisher's note} Springer Nature remains neutral with regard to jurisdictional claims in published maps and institutional affiliations.

\newpage
\begin{table*}[!htp]
\caption{\label{Tab1} Theories of Anomalous Hall effect for heavy-fermion systems.}
\begin{ruledtabular}
\begin{tabular}{ccc}
 Models &  $R_A\equiv \rho_{yx}^A/H$ & Notes\\ \hline
 Classic skew scattering \cite{Smit-1955,Smit1958} & $R_A\propto\chi\rho$ \\ \hline
 Resonant level \cite{Ramakrishnan-AHEKondo,Coleman-AHEMixed,Fert-HFAHE,Panwar-HFHall} & $R_A\propto \chi(1-\chi T)\rho$  &  $T>T_K$\\
                &  $R_A\propto \rho\chi$   &   $T_{coh}<T<T_K$\\ \hline
 Impurity-Anderson \cite{Fert-HFAHE}   & $R_A=\gamma_1\chi\rho$  & $T>T_K$
  \\
                                                         & $R_A=\gamma_2\chi\rho$ ($\gamma_2\neq\gamma_1$)  & $T_{coh}<T<T_K$   \\ \hline
  Periodic-Anderson \cite{Kohno-JMMM1990,Yamada-PTP1993,Kontani-AHEHF} &     $R_A\propto\chi\rho^0$ &    $T>T_{coh}$      \\
    & $R_A\propto\rho^2$          & $T<T_{coh}$   \\ \hline
  This work                       & $R_A\propto\chi\rho^0$   &    $T>T_{K}^{on}$ ($T_{K}^{on} \sim 3 T_{coh}$)  \\
    & -          &  $T<T_{K}^{on}$
\end{tabular}
\end{ruledtabular}
\end{table*}

\begin{table*}[!htp]
\caption{\label{Tab2} Magnetic space group (MSG), net magnetic moment $\mathbf{M}$, spatial inversion symmetry $\mathcal{P}$, the combined operation of inversion symmetry and time reversal symmetry $\mathcal{PT}$, and the topological feature of Ce$_3$TiSb$_5$ with different imposed magnetic configurations. The structures AFM-Zigzag$^{\text{F}}$ and AFM$^\text{F}_{60/120}$ respectively resemble $\mathbf{k}_1$ and $\mathbf{k}_2$ phases without $\mathbf{c}$-axis modulation. The abbreviations are: FM = Ferromagnetic; AFM = Antiferromagnetic; DSM = Dirac semimetal; WSM = Weyl semimetal; TDP = triply degenerate point. See more details in \textbf{SI}.}
\begin{ruledtabular}
\begin{tabular}{cccccccc}
Magnetic structures & Properties & MSG  & Net $\mathbf{M}$  & $\mathcal{P}$ & $\mathcal{PT}$ & Topology \\ \hline
FM$_z$ &    & P6$_3$/mc$^\prime$m$^\prime$ & $M_z$, $M_z$ & $+$  &  $-$  &  WSM  \\
AFM$_z$ &  & P6$_3$$^\prime$/mcm$^\prime$  & $0$, $0$ & $-$   &  $+$  &  Nodeless \\
AFM$^\text{F}_{120}$  & all-in-all-out &  P6$_3$$^\prime$/m$^\prime$c$^\prime$m & $0$, $0 $ & $+$  &  $-$  &  TDP-WSM \\
AFM$^\text{F}_{60/120}$ ($\sim \mathbf{k}_2$) & 2-in-1-out & Cmc$^\prime$m$^\prime$ & $M_y$, $M_y$  & $+$  &  $-$  &  WSM \\
AFM$^\text{AF}_{120}$ & all-in-all-out & P6$_3$/m$^\prime$cm   & $0$, $0$   & $-$   &  $+$  &  DSM\\
AFM$^\text{AF}_{60/120}$ & 2-in-1-out & Cmcm$^\prime$ & $0$, $0$   & $-$   &  $+$  &  Nodeless \\
AFM-Zigzag$^{\text{F}}$ ($\sim \mathbf{k}_1$) &  & Pm$^\prime$ma  &   $M_{x}$, $M_{x}$, $-M_{x}$, $-M_{x}$ & $-$ &  $+$       &  Nodeless   \\
La$_3$TiSb$_5$ & Non-magnetic  & P6$_3$/mcm  &   0, 0 & $+$ &  $+$     &  DSM   \\
\end{tabular}
\end{ruledtabular}
\end{table*}

\newpage
\begin{figure}[!htp]
\hspace*{-0pt}
\vspace*{-20pt}
\includegraphics[width=16.5cm]{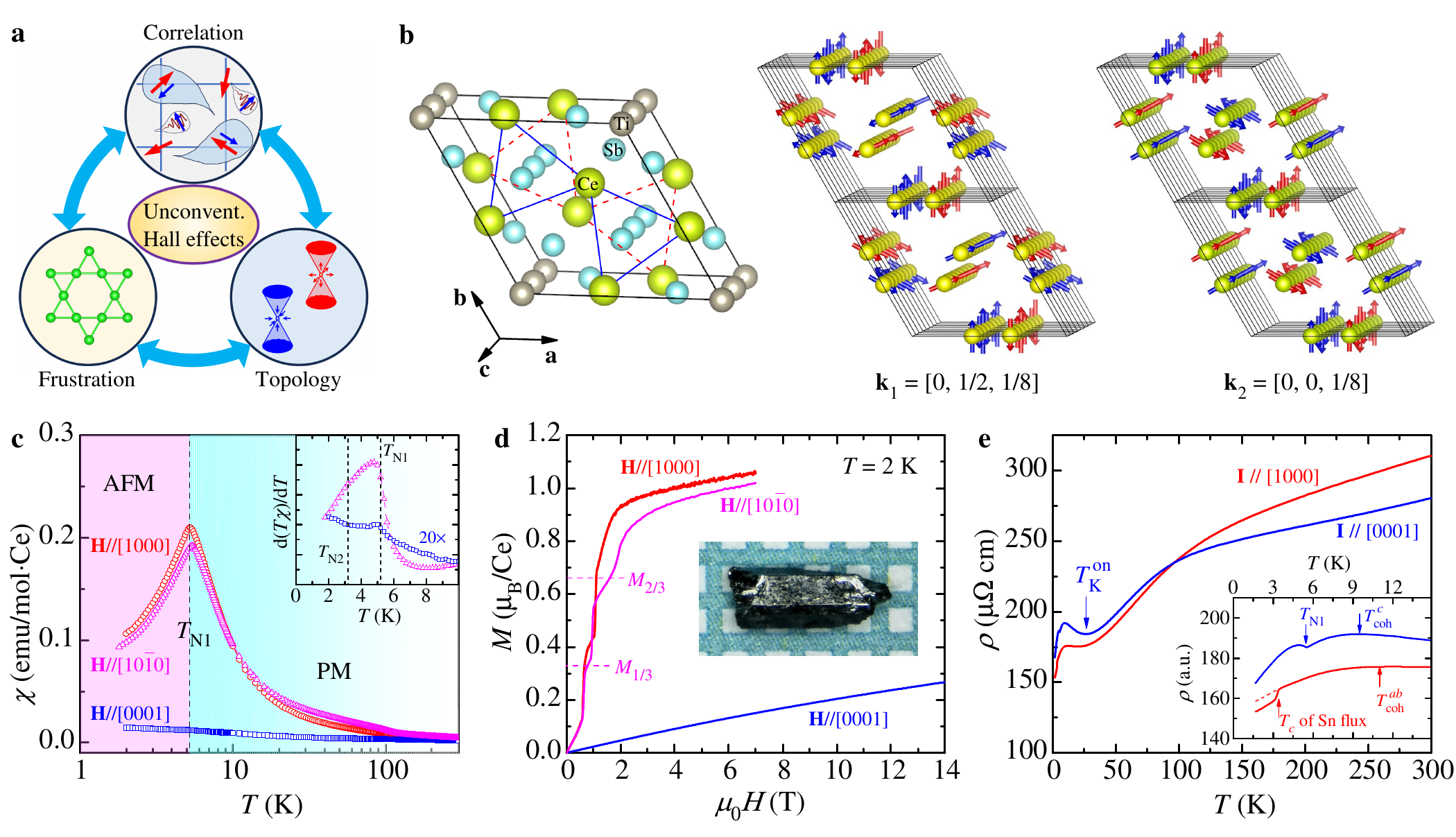}
\label{Fig1}
\end{figure}
\vspace*{-0pt}
\begin{spacing}{1.40}
\small{\textbf{Fig. 1 $|$ Crystalline structure, magnetic structure and sample characterization of Ce$_3$TiSb$_5$.} \textbf{a} A cartoon illustration of the interplay among electronic correlation, geometric frustration and topology in Kondo lattices, which can lead to unconventional Hall effects as exemplified by Ce$_3$TiSb$_5$. ``Correlation" panel: composite heavy quasiparticles in Kondo lattices arising from quantum entanglement of $f$ (red) and itinerant conduction (blue) electrons; ``Frustration" panel: kagome structure; ``Topology" panel: a pair of Weyl nodes with opposite chiralities. \textbf{b} Left, crystalline structure of Ce$_3$TiSb$_5$. Red and blue lines indicate Ce networks on $z$=0.25 and 0.75 planes, respectively, seeing also in Fig.~5\textbf{c}. Middle, magnetic structure with wave vector $\mathbf{k}_1=[0,~1/2,~1/8]$ for $T_{N2}<T<T_{N1}$. Only the Ce sublattice is shown for clarity. Right, magnetic structure with wave vector $\mathbf{k}_2=[0,~0,~1/8]$ for $T<T_{N2}$ \cite{Ritter-JPCM2021}. \textbf{c} Temperature dependence of magnetic susceptibility for $\mathbf{H}\parallel[1000]$ (red), $[10\bar{1}0]$ (green) and $[0001]$ (blue). Inset, $d(T\chi)/dT$ plots that exhibit both $T_{N1}$ and $T_{N2}$. \textbf{d} Field dependent magnetization at 2 K. $M(H)$ for $\mathbf{H}\parallel[1000]$ shows multiple steps hinting field-induced metamagnetic transition. $M_{1/3}$ and $M_{2/3}$ plateaus are visible for $\mathbf{H}\parallel[10\bar{1}0]$. $M$ is nearly linear with $H$ for [0001]. \textbf{e} Electrical resistivity as a function of $T$, measured with electrical current flowing along [1000] (red) and [0001] (blue) respectively. Onset of incoherent Kondo scattering ($T_K^{on}$) is signified by the minimum of $\rho(T)$. The inset displays a zoom-in plot for the low temperature region. A tiny trace of superconductivity is visible due to Sn-flux impurity, and can be suppressed by a small field 0.1 T. Kondo coherence ($T_{coh}$) establishes below $\sim$ 10 K for both orientations.}\\
\end{spacing}

\begin{figure*}[!htp]
\vspace*{0pt}
\hspace*{0pt}
\includegraphics[width=16.5cm]{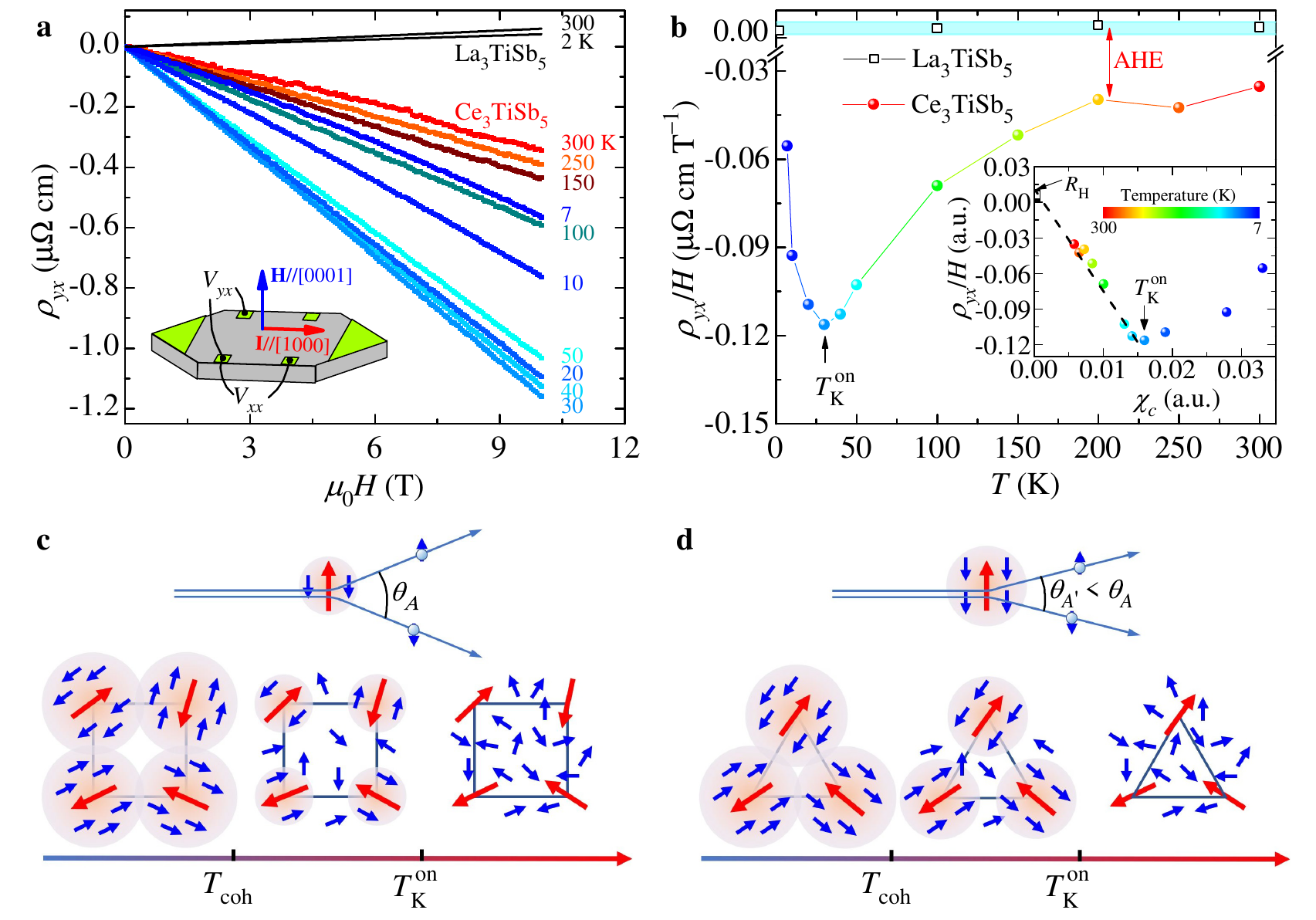}
\label{Fig2}
\end{figure*}
\vspace*{-20pt}
\small{\textbf{Fig. 2 $|$ Hall effect of Ce$_3$TiSb$_5$ in the paramagnetic state.} \textbf{a} Isothermal field dependent Hall resistivity $\rho_{yx}$ at different temperatures. Data of the non-magnetic reference La$_3$TiSb$_5$ are also shown for comparison. The inset depicts the schematic of transport measurements. \textbf{b} Hall resistivity divided by field, plotted as a function of $T$. The inset displays $\rho_{yx}/H$ vs. $\chi_c$ with $T$ as an implicit parameter. The linear fitting for $T>T_K^{on}$ leads to the normal Hall coefficient $R_H=+0.6(2)\times 10^{-10}$ m$^3$/C. The open squares represent the data of La$_3$TiSb$_5$. \textbf{c} Schematic diagrams of Kondo effects: non-screened regime for $T>T_{K}^{on}$, incoherent Kondo regime for $T_{coh}<T<T_{K}^{on}$, and coherent Kondo regime for $T<T_{coh}$. The hatched balls signify Kondo entanglement of $f$ (red) and conduction (blue) electrons. The upper panel sketches the skew scattering off a Kondo-screened local moment. \textbf{d} ibid, but in a geometrically frustrated lattice. Geometric frustration adds to the inherent frustration caused by Kondo effect, which further weakens the skew scattering in the frustration-promoted incoherent Kondo regime, viz $\theta_{A'}<\theta_{A}$.}\\

\newpage
\begin{figure*}[!htp]
\vspace*{-0pt}
\hspace*{-0pt}
\includegraphics[width=16.5cm]{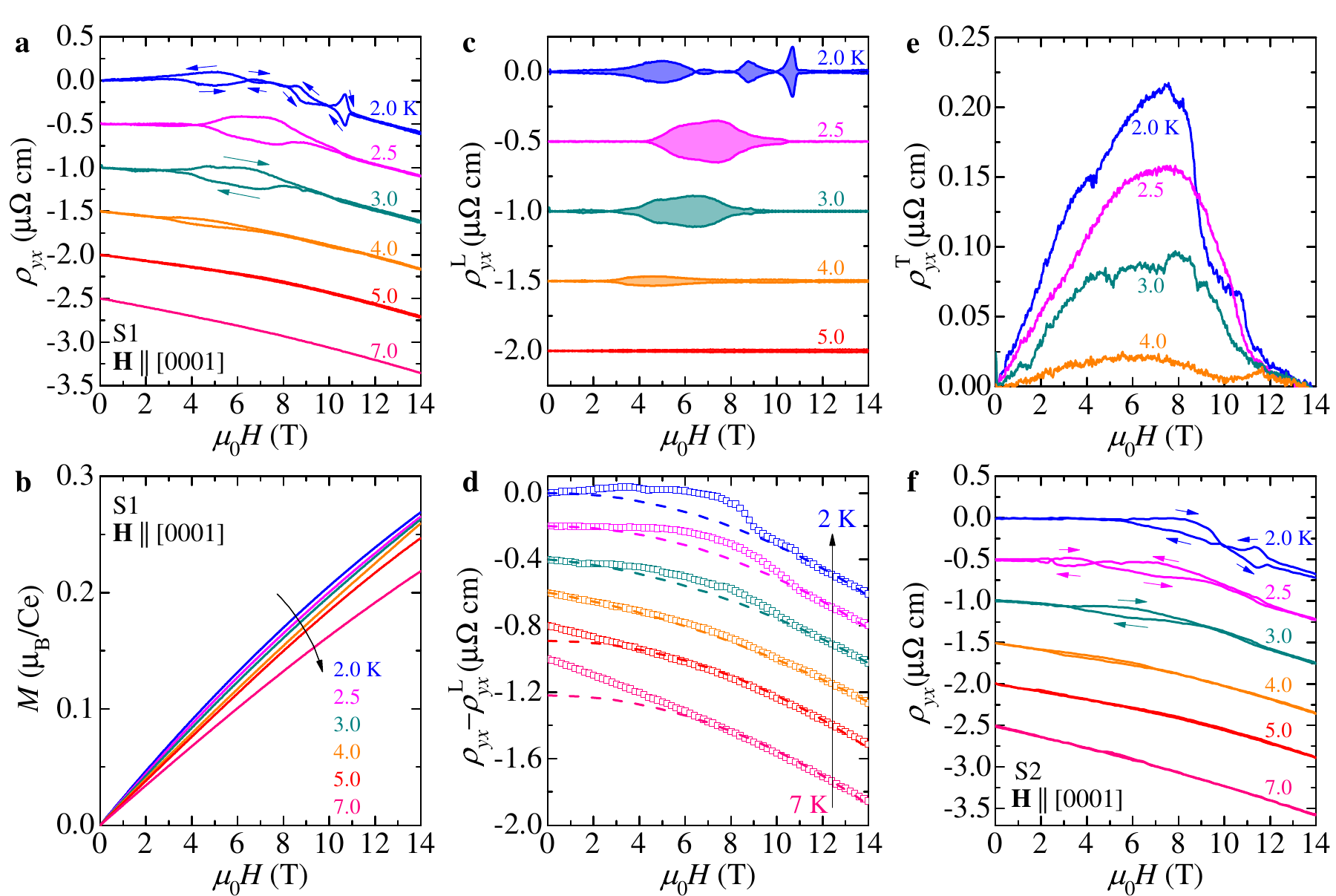}
\label{Fig3}
\end{figure*}
\vspace*{-20pt}
\small{\textbf{Fig. 3 $|$ Hall effect of Ce$_3$TiSb$_5$ in the magnetic ordered state.} \textbf{a} Total Hall resistivity as a function of $H$ near and below $T_{N1}$. $\rho_{yx}(H)$ shows clear history dependence. The curves are vertically shifted for clarity. The arrows indicate the direction of field sweep in the Hall effect measurements. \textbf{b} Field dependent magnetization is featureless at the same field and temperature. \textbf{c} Loop-shaped topological Hall resistivity $\rho_{yx}^L$, derived from $\rho_{yx}^{-+}-\rho_{yx}^{+-}$, where the superscripts ``$-+$" and ``$+-$" mean the direction of field sweep. \textbf{d} $\rho_{yx}-\rho_{yx}^L$ as a function of $H$ at various $T$. Note that for 7 and 5 K, $\rho_{yx}^L$ has no contribution, so $\rho_{yx}-\rho_{yx}^L$ is the same as $\rho_{yx}$. The dashed lines indicate the regime of $H^2$ dependence that moves to higher field as $T$ rises. \textbf{e} The extracted topological Hall resistivity $\rho_{yx}^T$. \textbf{f} Hall resistivity measured on a second Ce$_3$TiSb$_5$ sample (S2) grown in the same batch.} \\

\newpage
\begin{figure*}[!htp]
\vspace*{-0pt}
\hspace*{-0pt}
\includegraphics[width=15cm]{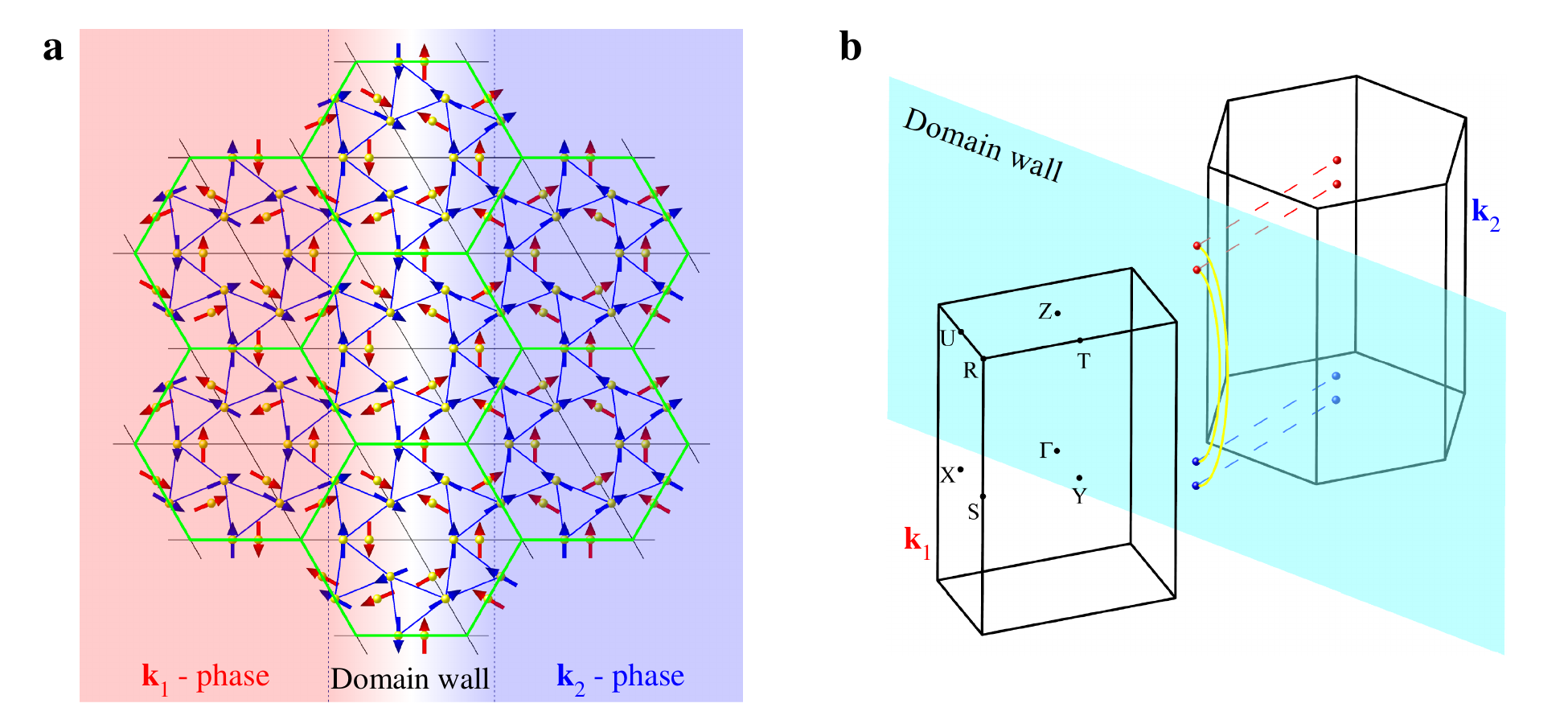}
\label{Fig4}
\end{figure*}
\vspace*{-0pt}
\small{\textbf{Fig. 4 $|$ Possible mechanisms for the loop-shaped Hall effect.} \textbf{a} Sketch of spin texture for domain wall between the $\mathbf{k}_1$ and $\mathbf{k}_2$ phases. The red and blue arrows stand for the Ce moments on different quasi-kagome networks. Abnormal history dependent real-space Berry curvature may be picked when carriers travel across this domain wall. \textbf{b} Chiral Fermi arc state (yellow lines) on the domain wall due to the Weyl points (red and blue balls) in the $\mathbf{k}_2$ phase. Such Fermi arcs provide a metallic interface that probably contributes an abnormal Hall voltage. The time-reversal invariant momenta (TRIM) points of the $\mathbf{k}_1$ phase are also depicted in this panel, while the TRIM points for the $\mathbf{k}_2$ phase are denoted in Fig.~5\textbf{a} for better clarity.}  \\

\newpage
\begin{figure*}[!htp]
\vspace*{-0pt}
\hspace*{-0pt}
\includegraphics[width=16.5cm]{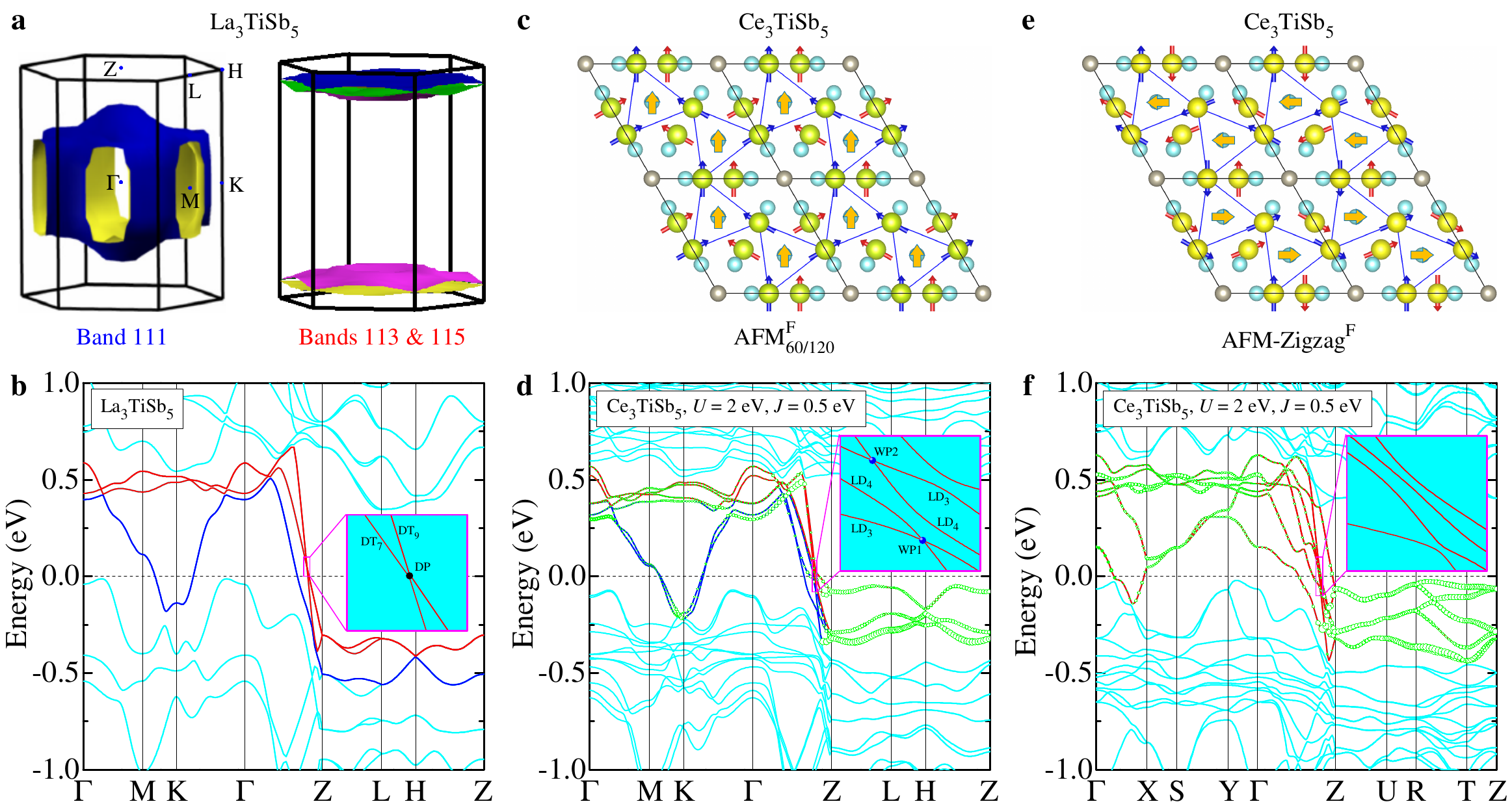}
\label{Fig5}
\end{figure*}
\vspace*{-20pt}
\small{\textbf{Fig. 5 $|$ Density-functional theory (DFT) calculated band structure of La$_{3}$TiSb$_5$ and Ce$_{3}$TiSb$_5$.} \textbf{a} Fermi-surface topology of La$_{3}$TiSb$_5$. The Fermi surface is constituted by a 3D-like pocket (Band 111) and two flat sheets (Bands 113 and 115). \textbf{b} Band structure of La$_{3}$TiSb$_5$. A type-\Rmnum{2} Dirac node is formed on $\Gamma$-Z by the crossing of bands 113 and 115. \textbf{c} AFM$^{\text{F}}_{60/120}$ magnetic structure that mimics the $\mathbf{k_2}$ phase of Ce$_3$TiSb$_5$. The moments of Ce on different layers are distinguished by colors (red and blue), while the net magnetic moment in a Ce$_6$ polyhegdron is marked by the orange arrows. \textbf{d} Band structure of Ce$_{3}$TiSb$_5$ in AFM$^{\text{F}}_{60/120}$ calculated with $U$ = 2 eV. The size of the green circles represents the weight of Ce-4$f$ orbitals. Two Weyl points are seen on $\Gamma$-$\text{Z}$, depicted by WP1 and WP2, respectively. \textbf{e} AFM-Zigzag$^\text{F}$ magnetic structure that mimics the $\mathbf{k_1}$ phase of Ce$_3$TiSb$_5$. \textbf{f} Band structure of Ce$_{3}$TiSb$_5$ in AFM-Zigzag$^\text{F}$. The crossing points on $\Gamma$-$\text{Z}$ near $E_F$ are gapped out, making it a trivial semimetal.} \\

\newpage
\begin{figure*}[!htp]
\vspace*{-0pt}
\hspace*{-0pt}
\includegraphics[width=9.5cm]{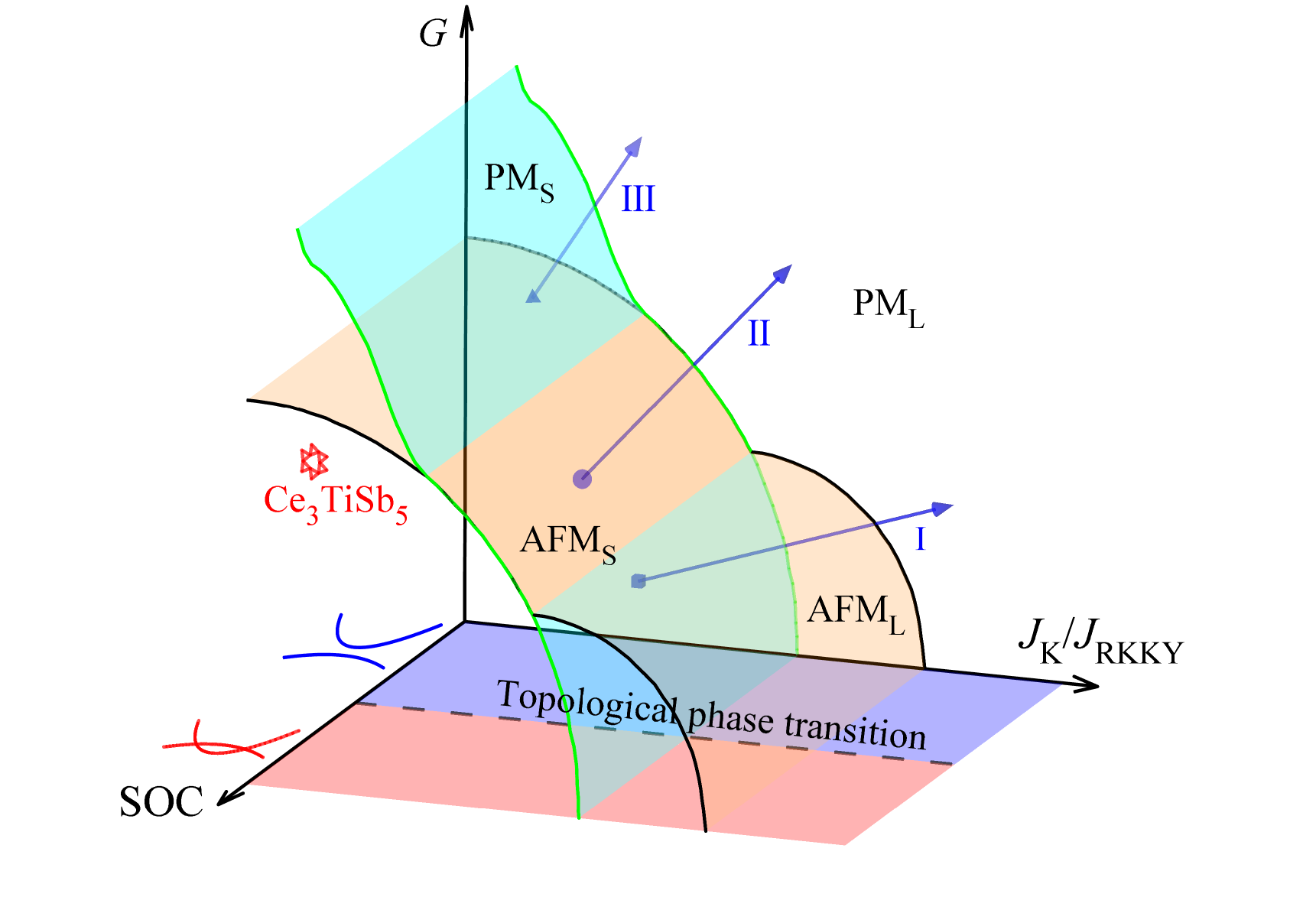}
\label{Fig6}
\end{figure*}
\vspace*{-20pt}
\small{\textbf{Fig. 6 $|$ Extended global phase diagram for heavy-fermion systems.} $G$ parameterizes frustration, $J_K$ characterizes Kondo coupling, and $J_{RKKY}$ signifies Ruderman-Kittel-Kasuya-Yosida interaction. A new degree of freedom, topology, that can be switched by spin-orbit coupling (SOC), is introduced to the classic global phase diagram \cite{Si-Global}. The abbreviations are: AFM = antiferromagnetic, and PM = paramagnetic; SDW = spin-density wave; QCP = quantum critical point. The subscript S (or L) denotes small (or large) Fermi surface. The blue arrows on the topologically trivial plane are the trajectories of quantum phase transition: I, SDW QCP ($\text{AFM}_\text{S}$-$\text{AFM}_\text{L}$-$\text{PM}_\text{L}$), e.g. CeRh$_{0.58}$Ir$_{0.42}$In$_5$ \cite{LuoY-CeRhIn5_IrQCP}; II, Kondo-breakdown QCP ($\text{AFM}_\text{S}$-$\text{PM}_\text{L}$), e.g. Ce(Cu,Au)$_6$ \cite{Lohneysen-CeCu6_AuQCP}, YbRh$_2$Si$_2$ \cite{Custers-YbRh2Si2QCP}, CeRhIn$_5$ \cite{Shishido-CeRhIn5dHvA}, and CeNiAsO \cite{LuoY-CeNiAsOQCP}; and III, quantum phase transition occurred in highly frustrated systems ($\text{AFM}_\text{S}$-$\text{PM}_\text{S}$-$\text{PM}_\text{L}$) as exemplified by CePdAl \cite{Zhao-CePdAlQCP}. Our results place Ce$_3$TiSb$_5$ at the highly-frustrated $\text{AFM}_\text{S}$ regime with non-trivial band topology.} \\


\newpage

\renewcommand{\thefigure}{S\arabic{figure}}
\renewcommand{\thetable}{S\arabic{table}}
\renewcommand{\theequation}{S\arabic{equation}}
\setcounter{table}{0}
\setcounter{figure}{0}
\setcounter{equation}{0}
\setcounter{page}{1}

\vspace{-15pt}

\begin{center}
\large
\textbf{Supplementary Information: } \\
\textbf{Rich unconventional Hall effects in a single quasi-kagome Kondo Weyl semimetal candidate Ce$_3$TiSb$_5$}\\
\small
\emph{}\\
Xiaobo He$^{1*}$, Ying Li$^{1,2*}$, Yongheng Ge$^{1*}$, Hai Zeng$^{1*}$, Shi-Jie Song$^{3}$, Jie Liu$^4$, Shuo Zou$^{1}$, Zhuo Wang$^{1}$, Yuke Li$^{4}$, Wenxin Ding$^{5}$, Jianhui Dai$^{4}$, Guang-Han Cao$^{3}$, Xiao-Xiao Zhang$^{1}$, Tianyou Zhai$^4$, Gang Xu$^{1\dag}$, and Yongkang Luo$^{1\ddag}$ \\

$^1$ {\it Wuhan National High Magnetic Field Center and School of Physics, Huazhong University of Science and Technology, Wuhan 430074, China;}\\
$^2$ {\it School of Science, Changzhou Institute of Technology, Changzhou 213032, China;}\\
$^3$ {\it School of Physics, Zhejiang University, Hangzhou 310058, China;}\\
$^4$ {\it State Key Laboratory of Materials Processing and Die \& Mould Technology, School of Materials Science and Engineering, Huazhong University of Science and Technology,Wuhan 430074, China;}\\
$^5$ {\it School of Physics and Hangzhou Key Laboratory of Quantum Matter, Hangzhou Normal University, Hangzhou 311121, China; and}\\
$^6$ {\it School of Physics and Optoelectronics Engineering, Anhui University, Hefei, 230601, China}\\

\end{center}

\normalsize

In this \textbf{Supplementary Information (SI)}, we provide the additional results that further support the discussions in the main text, including Energy-dispersive X-ray spectroscopy (EDS), Wavelength-dispersive X-ray (WDX) spectroscopy, Laue X-ray diffraction, single-crystal X-ray diffraction, the mechanism of AHE, MFM images for magnetic domains, and DFT calculations.\\

\newpage

\textbf{SI {I}: A\lowercase{dditional} \lowercase{sample characterization}}

\begin{figure*}[!htp]
	\vspace*{-0pt}
	\includegraphics[width=15cm]{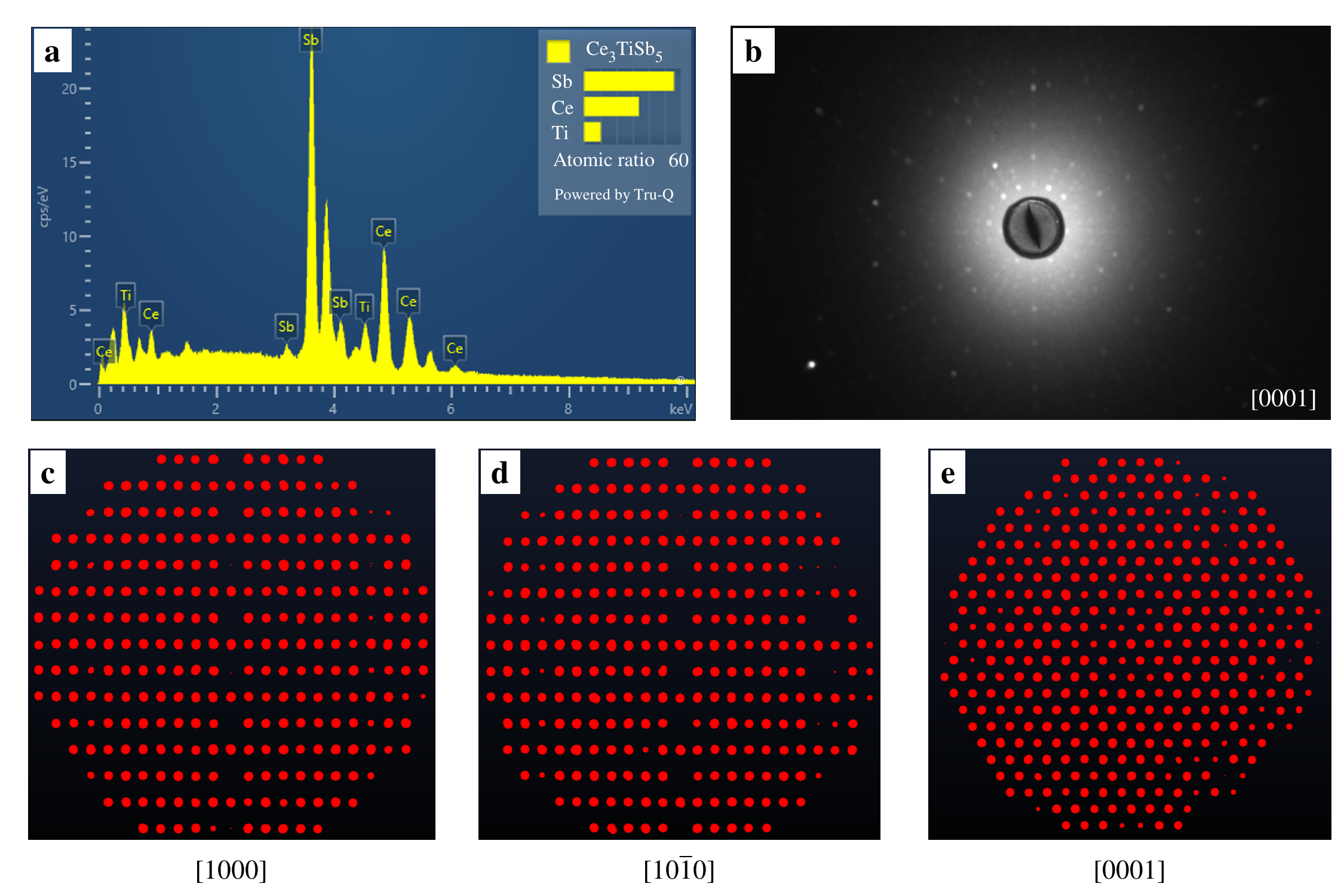}
	\label{Fig.S1}
\vspace*{-5pt}
\small
\begin{flushleft}
\justifying{
    \textbf{Supplementary Figure 1 $|$ Sample characterization of Ce$_3$TiSb$_5$.} \textbf{a} EDS results, which indicate the atomic ratio Ce : Ti : Sb $\approx$ 33.4(4) at.\% : 10.8(2) at.\% : 55.8(5) at.\%. \textbf{b} Laue X-ray diffraction. The nice pattern on the [0001] surface guarantees the good crystallization and nearly-perfect orientation. \textbf{c Single-crystal X-ray diffraction on [1 0 0 0] plane. \textbf{d} [1 0 $\bar{1}$ 0] plane. \textbf{e} [0 0 0 1] plane. }
    }
\end{flushleft}
\normalsize
\end{figure*}

\begin{figure*}[!htp]
	\vspace*{-10pt}
	\includegraphics[width=15cm]{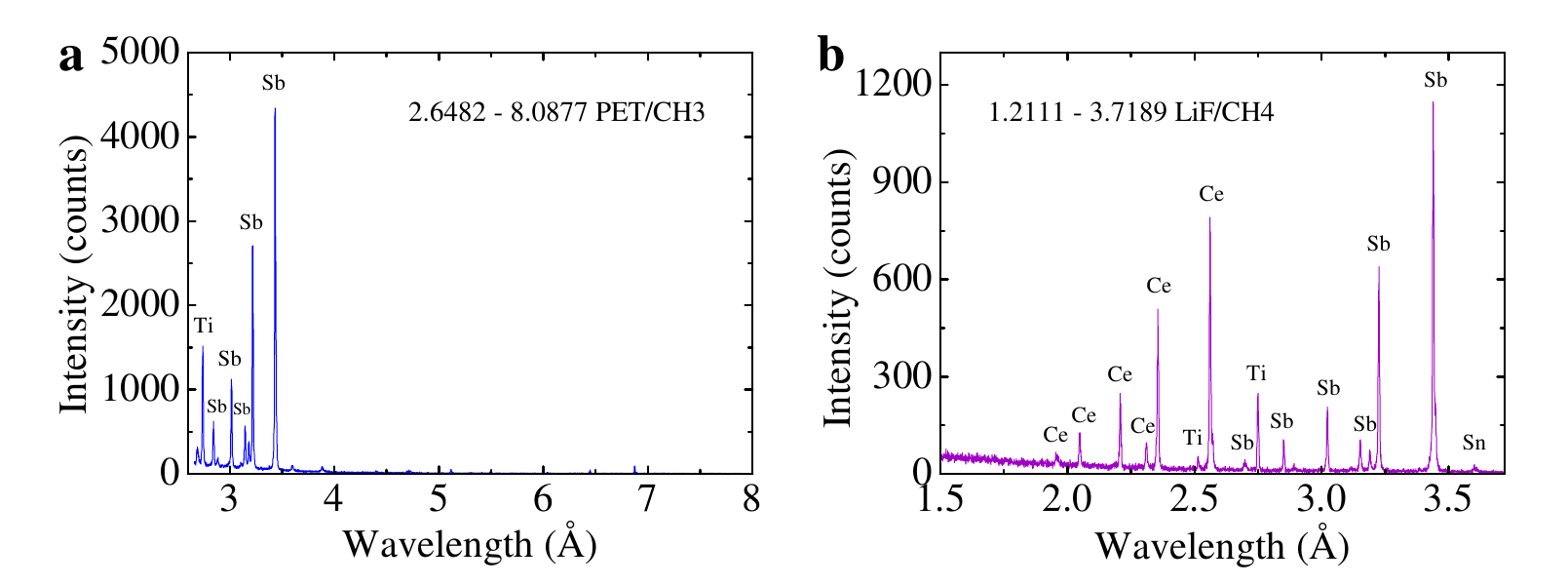}
	\label{Fig.S2}
\vspace*{-10pt}
\small
\begin{flushleft}
\justifying{
    \textbf{Supplementary Figure 2 $|$ Wavelength-dispersive X-ray (WDX) spectra of Ce$_3$TiSb$_5$.} A quantitative analysis leads to the molar ratio Ce : Ti : Sb $\approx$ 32.06(7) at.\% : 10.58(3) at.\% : 56.59(8) at.\%, close to the stoichiometry of Ce$_3$TiSb$_5$. A small amount of Sn-flux impurity ($\sim 0.77$ at.\%) was found in the measurement.
    }
\end{flushleft}
\normalsize
\end{figure*}

\begin{table}[!ht]
\begin{flushleft}
\justifying{
Supplementary Table 1: Single-crystal XRD refinement for Ce$_3$TiSb$_5$ at 296 K. Measurements with Cu $K_\alpha$ radiation (1.54184 \AA). Refinement software: SHELXL-2018/3.}
\end{flushleft}
\vspace*{-33pt}
\begin{center}
\def\temptablewidth{1\columnwidth}
{\rule{\temptablewidth}{1pt}}
\begin{tabular*}{\temptablewidth}{@{\extracolsep{\fill}}cc}
Composition ~~~~~~~~~~~~~~~~~~~~~~~~~~~~~~~~~~~         &  Ce$_3$TiSb$_5$           \\
Formula weight (g/mol) ~~~~~~~~~~~~~~~~~~~~      &  1077.02                      \\
Crystal system ~~~~~~~~~~~~~~~~~~~~~~~~~~~~~~~~      & Hexagonal                 \\
Space group ~~~~~~~~~~~~~~~~~~~~~~~~~~~~~~~~~~~         &  $P6_3/mcm$ (No. 193)             \\
$a$ (\AA) ~~~~~~~~~~~~~~~~~~~~~~~~~~~~~~~~~~~~~~~~~~~           &  9.4281(1)     \\
$c$ (\AA)  ~~~~~~~~~~~~~~~~~~~~~~~~~~~~~~~~~~~~~~~~~~~          &  6.2320(5)   \\
$V$ (\AA$^{3}$) ~~~~~~~~~~~~~~~~~~~~~~~~~~~~~~~~~~~~~~~~~~     &  479.737(6)    \\
$Z$  ~~~~~~~~~~~~~~~~~~~~~~~~~~~~~~~~~~~~~~~~~~~~~~~~~          &  4   \\
$\rho$ (g/cm$^{3}$) ~~~~~~~~~~~~~~~~~~~~~~~~~~~~~~~~~~~~~~ &  10.822                     \\
Absorption coefficient (cm$^{-1}$)  ~~~~~~~~~~~~~~ &  11.722               \\
$2\theta$ range ~~~~~~~~~~~~~~~~~~~~~~~~~~~~~~~~~~~~~~~~     &  5.4178 - 74.255$^{\circ}$         \\
$h, k, l$ range ~~~~~~~~~~~~~~~~~~~~~~~~~~~~~~~~~~~~ & $-11\leq h\leq 11$, $-11\leq k\leq 11$, $-7\leq l\leq 7$  \\
Collected reflections ~~~~~~~~~~~~~~~~~~~~~~~~~  &   9007                           \\
Independent reflections ~~~~~~~~~~~~~~~~~~~~~ &   858                                   \\
Goodness of fit on $F^2$ ~~~~~~~~~~~~~~~~~~~~~~~ &   1.386                                   \\
$R_1^{\dag}$, $w R_2^{\ddag}$ [$I>2\delta(I)$] ~~~~~~~~~~~~~~~~~~~~~~~~~~~&   3.32\%, 5.86\%    \\
\end{tabular*}
{\rule{\temptablewidth}{1pt}}
\end{center}
\small
\vspace*{-30pt}
\begin{flushleft}
\justifying{
$^{\dag}$ $R_1=\sum||F_{obs}|-|F_{cal}||/\sum|F_{obs}|$.\\
$^{\ddag}$ $w R_2=[\sum w(F_{obs}^2-F_{cal}^2)^2/\sum w (F_{obs}^2)^2]^{1/2}$, \\
$~~~~w=1/[\sigma^2F_{obs}^2+(a\cdot P)^2+b\cdot P]$, where  $P=[\mathrm{max}(F_{obs}^2)+2F_{cal}^2]/3$.
}
\end{flushleft}
\normalsize
\end{table}

\begin{table}[!ht]
\begin{flushleft}
\justifying{
Supplementary Table 2: Atomic coordinates and equivalent isotropic displacement parameters $U_{eq}$ of Ce$_3$TiSb$_5$. $U_{eq}$ is taken as 1/3 of the trace of the orthogonalized $U_{ij}$ tensor.}
\end{flushleft}
\vspace*{-33pt}
\begin{center}
\def\temptablewidth{1\columnwidth}
{\rule{\temptablewidth}{1pt}}
\begin{tabular*}{\temptablewidth}{@{\extracolsep{\fill}}ccccccc}
Atoms   &  Wyck.  &  $x$          &     $y$      &     $z$          & $U_{eq}$ (\AA$^2$) \\\hline
  Ce     &   6g    &  0.6173(1)   &  1  &    3/4           &   0.00195(9)            \\
  Ti    &   2b    &  1   &  1  &    1/2    &      0.00155(2)            \\
  Sb1    &   6g    &  1   &  0.25338(9)  &    3/4            &   0.00202(1)            \\
  Sb2    &   4d    &  1/3   &  2/3  &    1/2           &   0.00189(9)            \\
\end{tabular*}
{\rule{\temptablewidth}{1pt}}
\end{center}
\end{table}

\clearpage
\textbf{SI {II}: A\lowercase{dditional} data of unconventional Hall effects}

The origin of AHE is usually classified into three mechanisms: skew scattering, side jump, and intrinsic mechanism connected to Berry curvature \cite{Nagaosa-AHERMP}. The mechanism of skew scattering can be distinguished from the other two by the exponent $\beta$ in $\rho_{yx}^A(T,H) \propto \rho^{\beta}M(T,H)$. In incoherent Kondo regime of heavy-fermion metals, previous experimental and theoretical works revealed a modified skew scattering scenario with $\beta\approx 0$ \cite{Kohno-JMMM1990,Yamada-PTP1993,Kontani-AHEHF}, and this is partly confirmed in our data for $T>T_K^{on}$. Here we show how standard skew scattering ($\beta\approx 1$), intrinsic and side-jump ($\beta\approx 2$) fail in our results. Supplementary Figure 3(a) and (b) respectively display $\rho_{yx}/H$ vs. $\rho \chi_c$ and $\rho_{yx}/H$ vs. $\rho^2 \chi_c$ plots with $T$ an implicit parameter. The data points of the non-magnetic reference La$_3$TiSb$_5$ are also shown as the case in the limit of $\chi_c \rightarrow 0$. One clearly sees that neither $\beta = 1$ nor 2 works for Ce$_3$TiSb$_5$.

\begin{figure*}[!htp]
	\vspace*{10pt}
	\includegraphics[width=\textwidth]{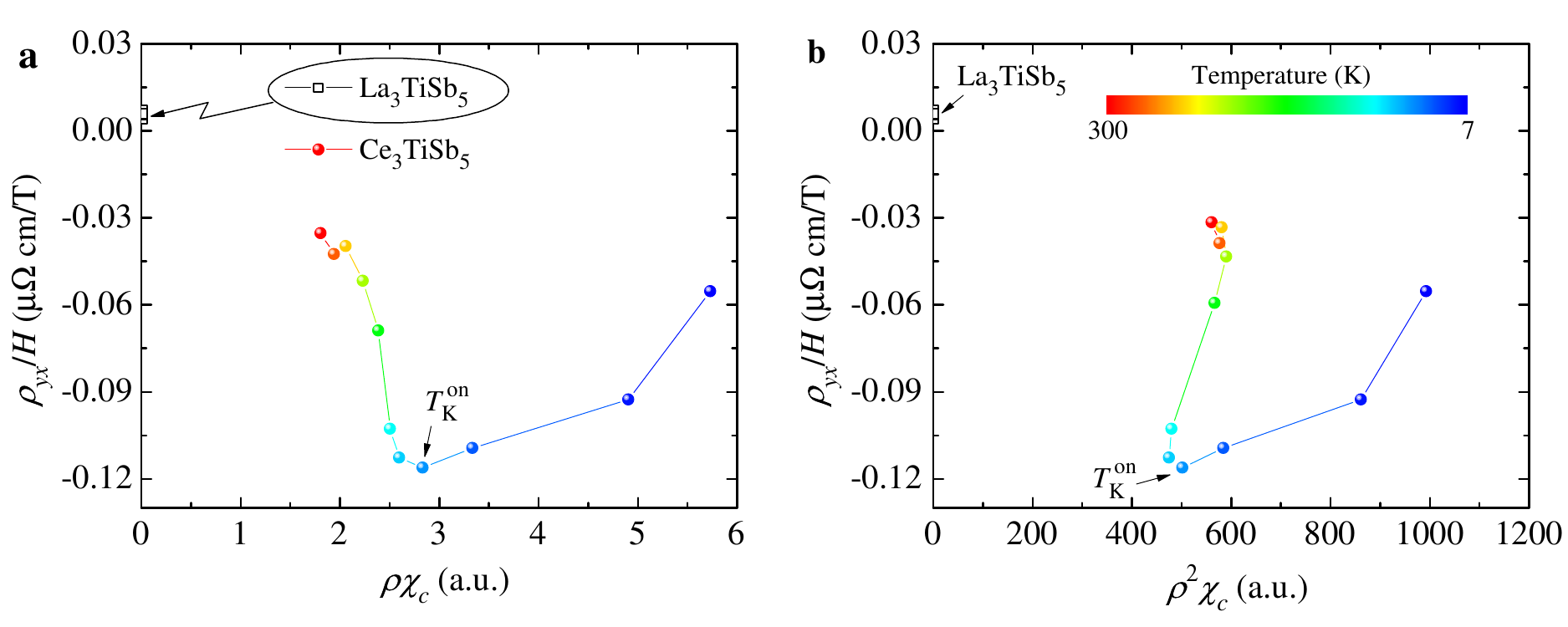}
	\label{Fig.S3}
\vspace*{-20pt}
\small
\begin{flushleft}
\justifying{
    \textbf{Supplementary Figure 3 $|$ Additional AHE analysis.} \textbf{a} $\rho_{yx}/H$ vs. $\rho\chi_c$ (conventional skew scattering). \textbf{b} $\rho_{yx}/H$ vs. $\rho^2\chi_c$ (side-jump or intrinsic).
    }
\end{flushleft}
\normalsize
\end{figure*}

\begin{figure*}[!htp]
	\vspace*{10pt}
	\includegraphics[width=0.6\textwidth]{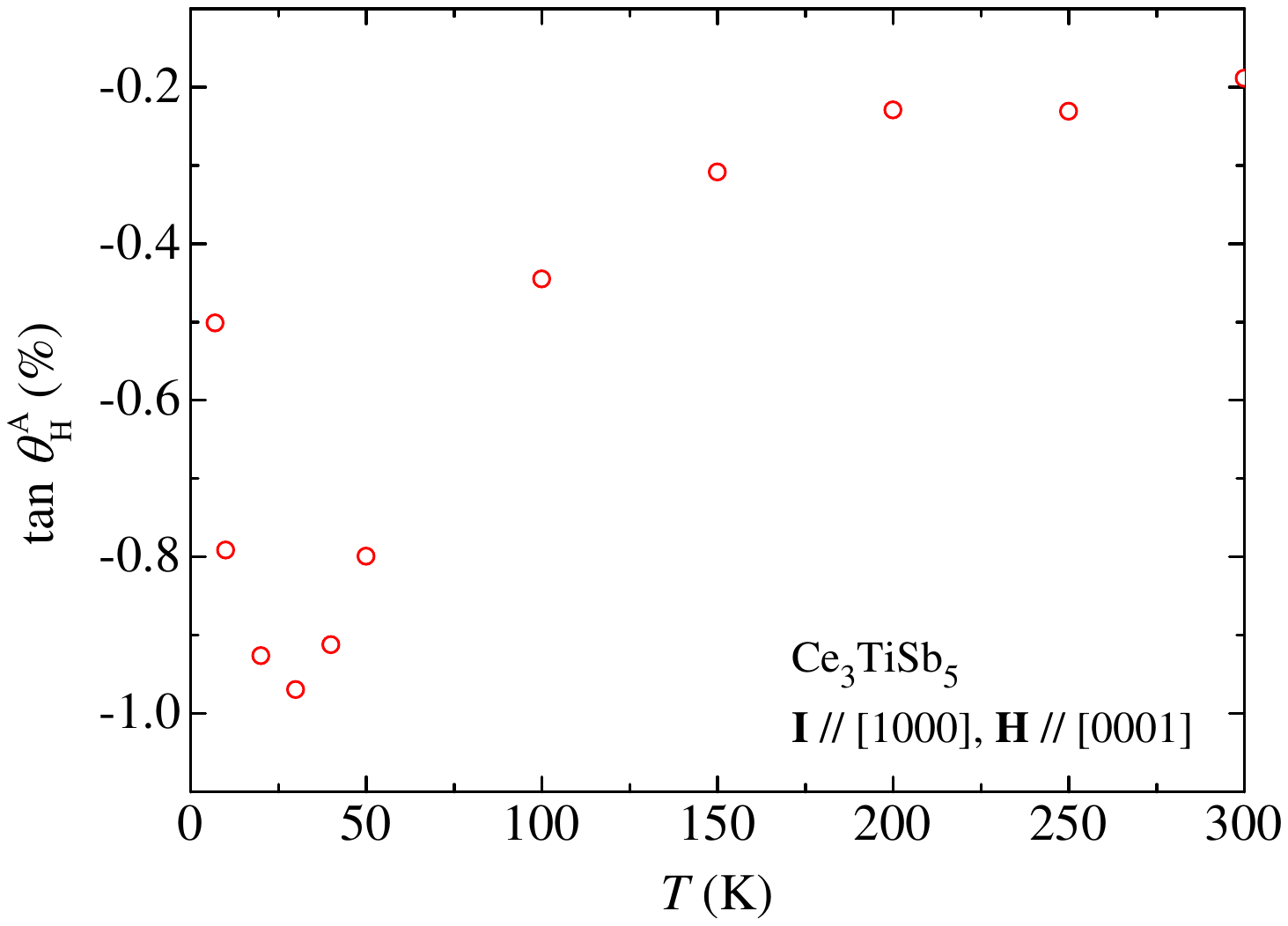}
	\label{Fig.S4}
\vspace*{-10pt}
\small
\begin{flushleft}
\justifying{
    \textbf{Supplementary Figure 4 $|$ Anomalous Hall angle of Ce$_3$TiSb$_5$ as a function of temperature.} $\tan{\theta_H^A}=\rho_{yx}^A/\rho_{xx}$.
    }
\end{flushleft}
\normalsize
\end{figure*}

\begin{figure*}[!htp]
	\vspace*{10pt}
	\includegraphics[width=\textwidth]{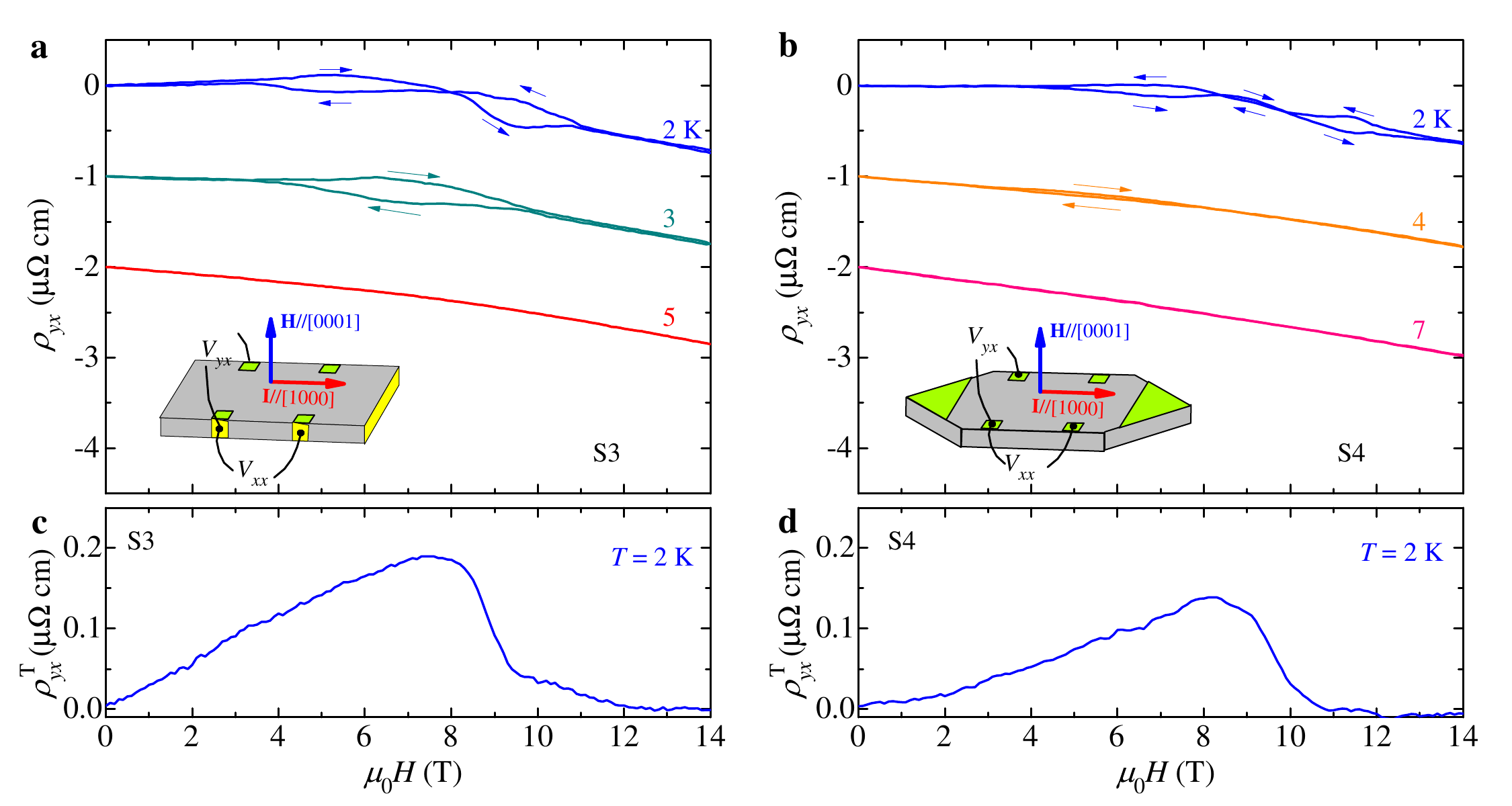}
	\label{Fig.S5}
\vspace*{-20pt}
\small
\begin{flushleft}
\justifying{
    \textbf{Supplementary Figure 5 $|$ Additional Hall resistivity of Ce$_3$TiSb$_5$ for below $T_{N1}$, measured on S3 and S4.} The contacts on S4 were prepared similarly as those of S1 and S2, while sample S3 was polished into a rectangular shape, and the current ends were fully covered with silver paint. For all these measurements, the electric current was applied along [1000], and the magnetic field was along [0001]. Despite the different contact configurations and loop structures, the obtained THE components for the two samples appear to be similar.
    }
\end{flushleft}
\normalsize
\end{figure*}

\clearpage
\textbf{SI III: M\lowercase{icroscopic image for domains}}

\begin{figure*}[!htp]
	\vspace*{0pt}
	\includegraphics[width=0.7\textwidth]{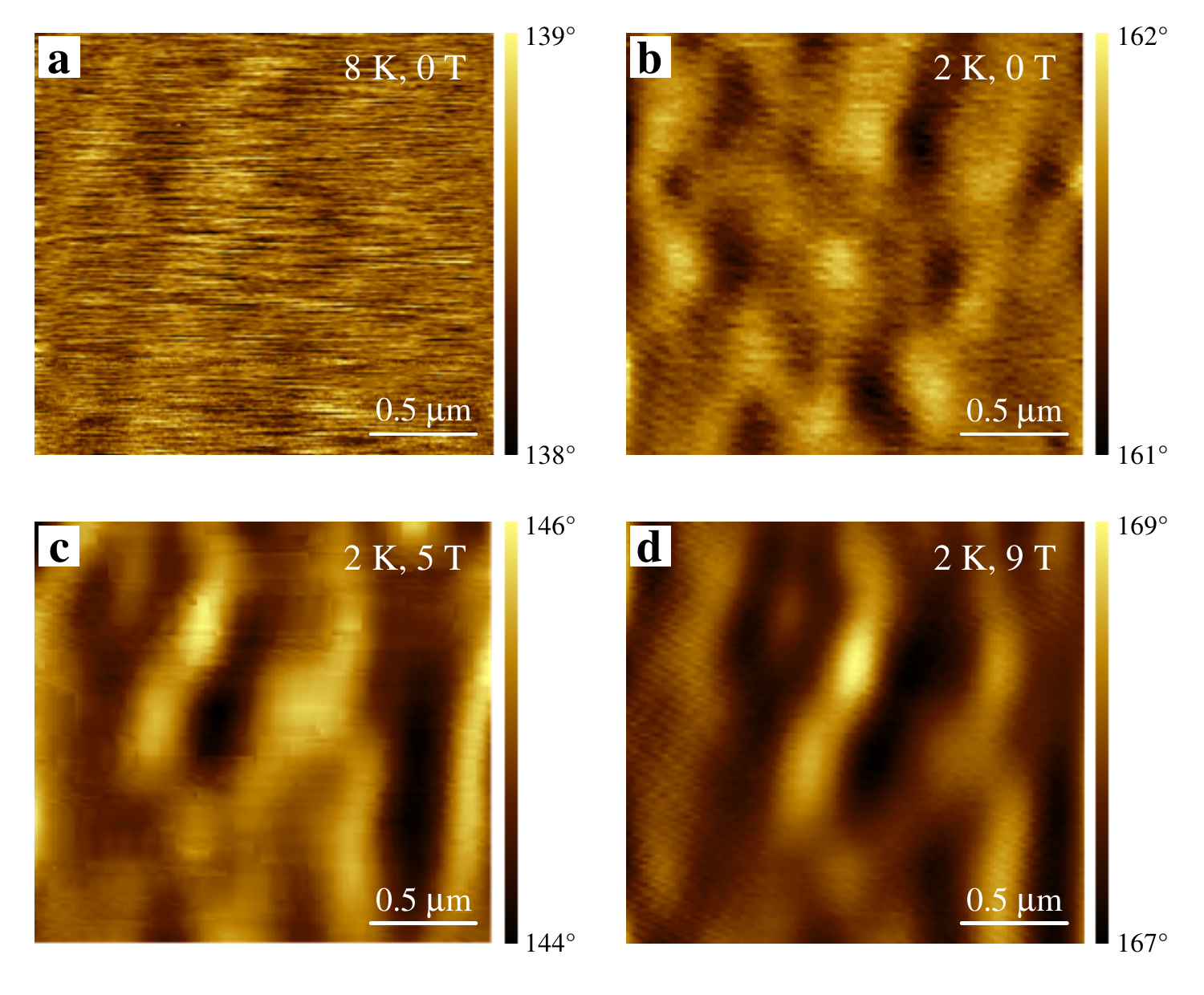}
	\label{Fig.S6}
\vspace*{-0pt}
\small
\begin{flushleft}
\justifying{
    \textbf{Supplementary Figure 6 $|$ Magnetic-Force Microscopy (MFM) images of Ce$_3$TiSb$_5$ on a fine-polished [0001] surface.} \textbf{a} At 8 K (above $T_{N1}$), no sizable magnetic domains can be identified. \textbf{b} At 2 K, magnetic domains are clearly visible. These domains evolve under magnetic field $\mathbf{H}\parallel[0001]$. \textbf{c} $\mu_0H = 5$ T. \textbf{d} $\mu_0H = 9$ T.
    }
\end{flushleft}
\normalsize
\end{figure*}

\clearpage
\textbf{SI IV: A\lowercase{dditional} DFT \lowercase{calculations}}

\begin{table}[!ht]
\begin{flushleft}
\justifying{
Supplementary Table 3: Comparison of different magnetic configurations for Ce$_3$TiSb$_5$. The magnetic structures of these configurations are depicted in Supplementary Figure 7 The energies are with respect to the AFM$^{\text{F}}_{60/120}$ ground state. Calculations made with $U=2$ eV and $J=0.5$ eV.}
\end{flushleft}
\vspace*{-33pt}
\begin{center}
\def\temptablewidth{1\columnwidth}
{\rule{\temptablewidth}{1pt}}
\begin{tabular*}{\temptablewidth}{@{\extracolsep{\fill}}cccccccc} \hline
     & FM$_z$  &  AFM$_z$  &  AFM$^{\text{F}}_{120}$  &  AFM$^{\text{AF}}_{120}$  &  AFM$^{\text{F}}_{60/120}$  & AFM$^{\text{AF}}_{60/120}$ & AFM-Zigzag$^{\text{F}}$\\ \hline
$E$ (meV/Ce)  &  1.440     &  2.087     &  1.930   &  3.678  &    0      &  4.303     &  11.979\\ \hline
\end{tabular*}
{\rule{\temptablewidth}{1pt}}
\end{center}
\normalsize
\end{table}

\begin{figure*}[!htp]
	\vspace*{-34pt}
	\includegraphics[width=16.5cm]{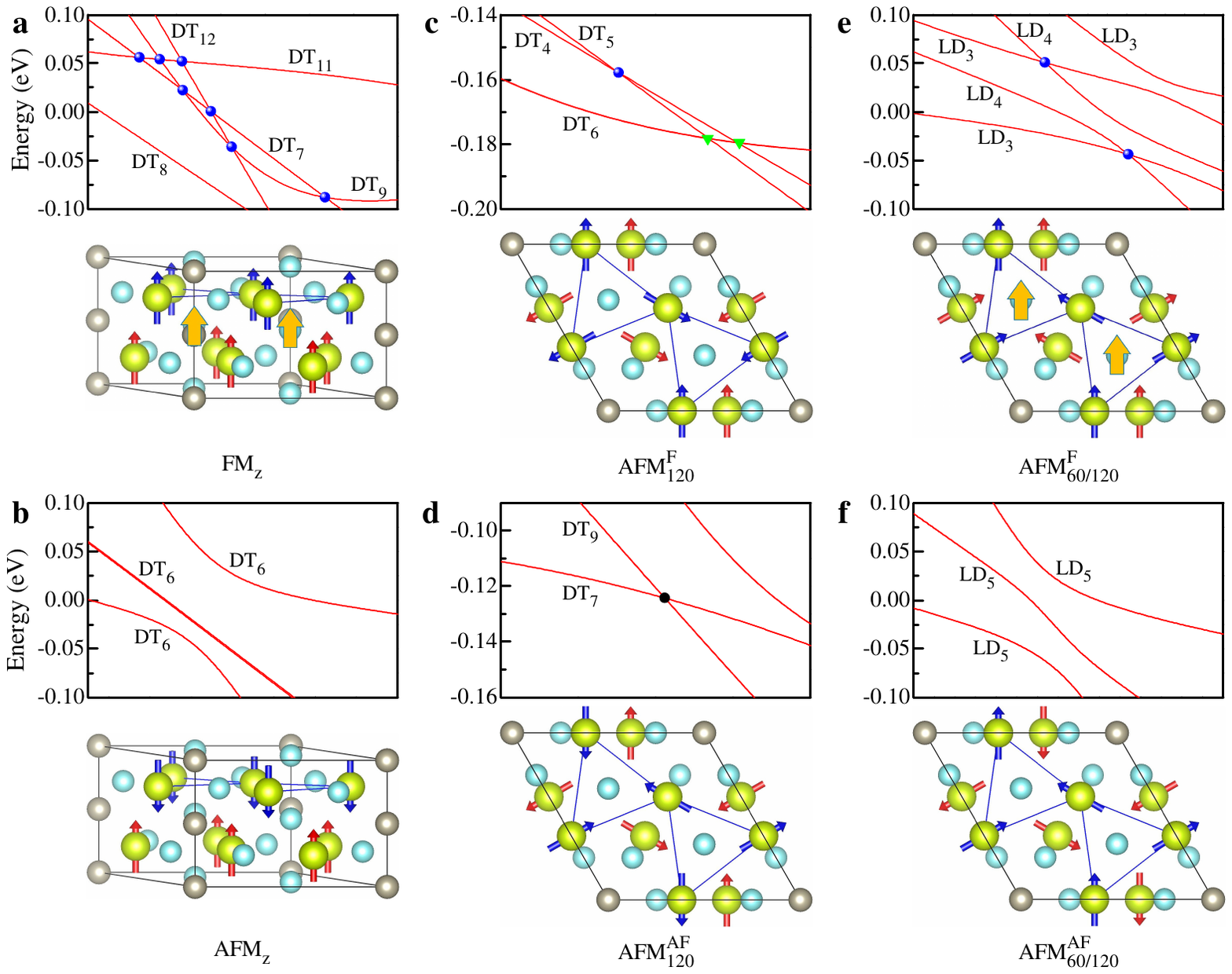}
	\label{Fig.S7}
\vspace*{-20pt}
\small
\begin{flushleft}
\justifying{
    \textbf{Supplementary Figure 7 $|$ Topological states of Ce$_3$TiSb$_5$ for different magnetic states.} Enlarged local band structure along $\Gamma$-$\text{Z}$ to show the topological features for \textbf{a} FM$_z$, \textbf{b} AFM$_z$, \textbf{c} AFM$^{\text{F}}_{120}$, \textbf{d} AFM$^{\text{AF}}_{120}$, \textbf{e} AFM$^{\text{F}}_{60/120}$, and \textbf{f} AFM$^{\text{AF}}_{60/120}$. The calculations were made with $U$ = 2 eV and $J$ = 0.5 eV. The superscripts ``F" (or ``AF") means ferromagnetic (or antiferromagnetic) inter-sublattice coupling. For AFM$^{\text{F}}_{120}$ and AFM$^{\text{AF}}_{120}$, the Ce moments are aligned within the $\mathbf{ab}$ plane, and the angles spanned by the nearest neighbours are all 120 $^{\circ}$; whereas for AFM$^{\text{F}}_{60/120}$ and AFM$^{\text{AF}}_{60/120}$, the angles are either 60 $^{\circ}$ or 120 $^{\circ}$. For AFM$^{\text{AF}}_{120}$, the Dirac point preserves (black dots); for FM$_z$ and  AFM$^{\text{F}}_{60/120}$, the Dirac point splits into Weyl points (blue balls); while in the case of AFM$^{\text{F}}_{120}$, besides the Weyl point at the crossing between DT$_4$ and DT$_5$, two triply degenerated points emerge as shown by the green triangles because of the doubly degenerated DT$_6$ band. No Dirac or Weyl point is observed near Fermi energy in AFM$_z$ and AFM$^{\text{AF}}_{60/120}$.
    }
\end{flushleft}
\normalsize
\end{figure*}

\begin{figure*}[!htp]
	\vspace*{10pt}
	\includegraphics[width=\textwidth]{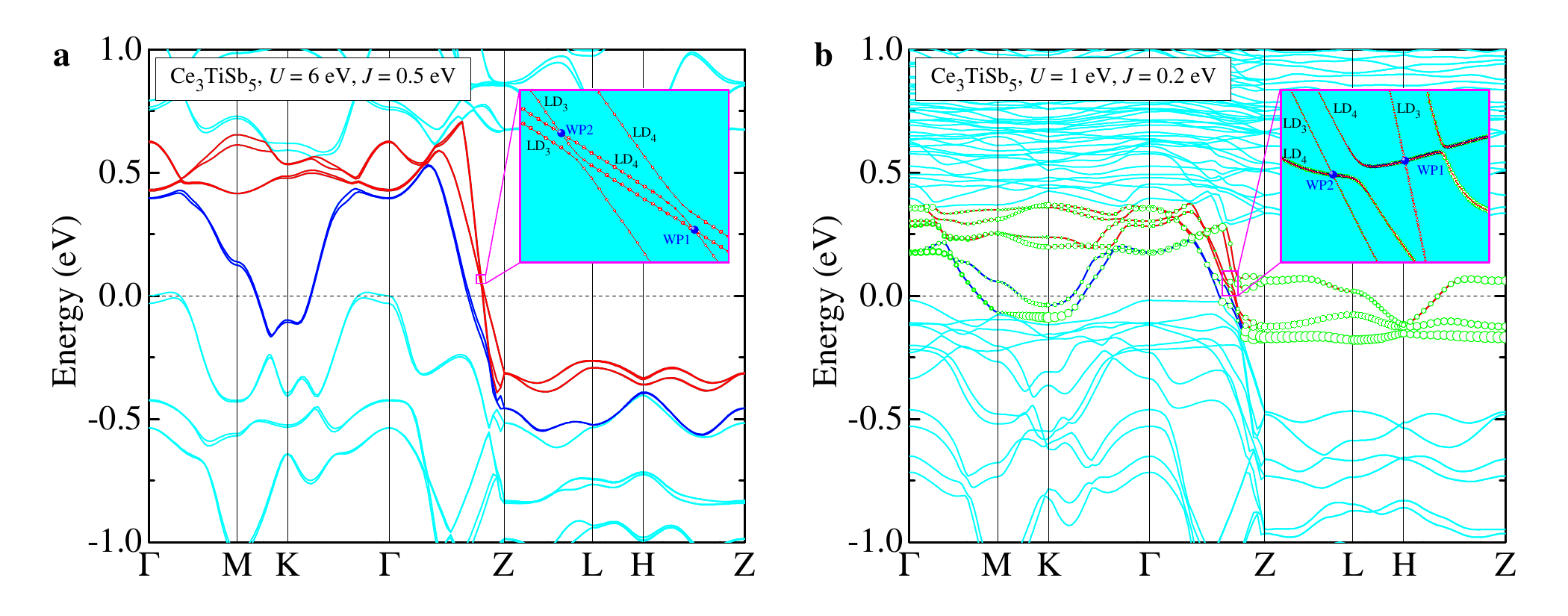}
	\label{Fig.S8}
\vspace*{-10pt}
\small
\begin{flushleft}
\justifying{
    \textbf{Supplementary Figure 8 $|$ DFT calculations of Ce$_3$TiSb$_5$ for magnetic state AFM$^\text{F}_{60/120}$ with different $U$ and $J$.} \textbf{a} $U=6$ eV and $J=0.5$ eV. The Ce-$4f$ orbitals are fully localized and pushed away from the Fermi level. The crossing bands near the $E_F$ are mainly derived from the $p_{x,y}$ and $p_z$ orbitals of the Sb1 atoms (located within the Ce layers). \textbf{b} DFT calculations with $U=1$ eV and $J=0.2$ eV. The size of the green circles represents the weight of Ce-4$f$ orbitals. The narrow Ce-$4f$ bands are close to the Fermi level. As a result, there is significant component mixing among Ce-$4f$ (green), Ti-$3d$ (navy), and Sb1-$5p$ (red) orbitals. For both cases, symmetry-protected Weyl points WP1 and WP2 are observed on $\Gamma$-$\text{Z}$ at the crossing of the bands with distinct representations LD$_3$ and LD$_4$.
    }
\end{flushleft}
\normalsize
\end{figure*}

\clearpage
\textbf{SI V: C\lowercase{omparison} between Ce$_3$TiSb$_5$ and CePdAl}

\begin{figure*}[!htp]
	\vspace*{10pt}
	\includegraphics[width=\textwidth]{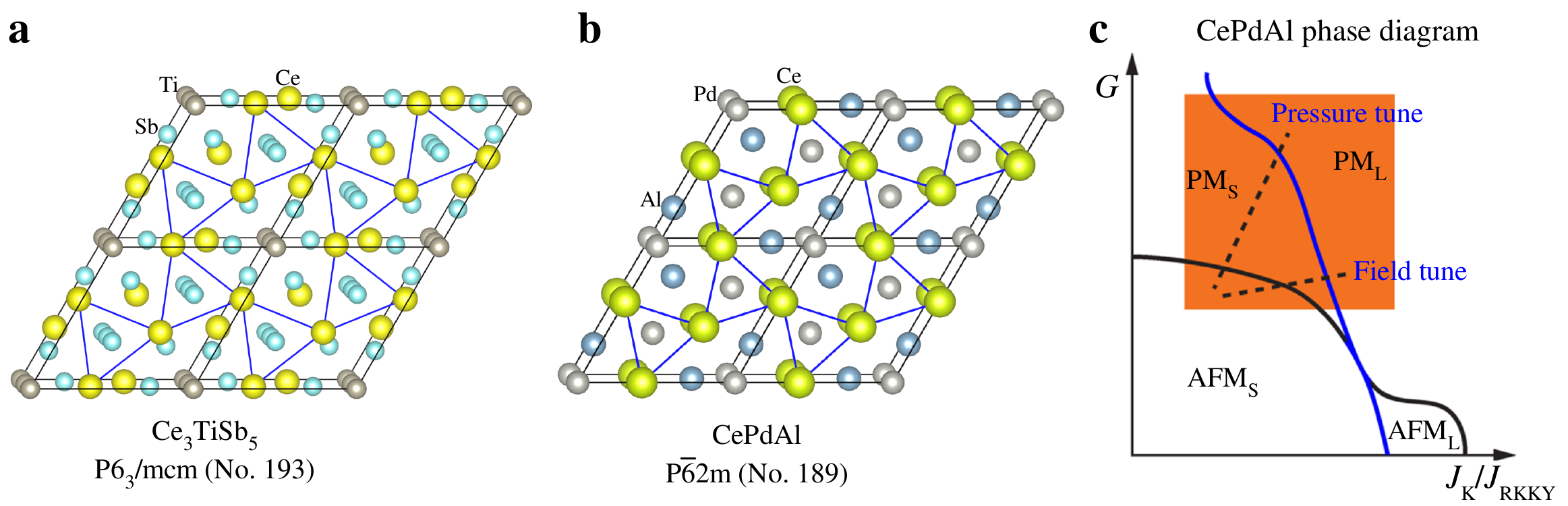}
	\label{Fig.S9}
\vspace*{-10pt}
\small
\begin{flushleft}
\justifying{
    \textbf{Supplementary Figure 9 $|$ Comparison between Ce$_3$TiSb$_5$ and CePdAl.} \textbf{a} Crystalline structure of Ce$_3$TiSb$_5$. \textbf{b} Crystalline structure of CePdAl. Note that the kagome spin-ice material HoAgGe shares the same crystalline structure with CePdAl \cite{Zhao-HoAgGe}. \textbf{c} Phase diagram of CePdAl tuned by pressure and field, reproduced from \cite{Zhao-CePdAlQCP}. For both tunings, the system undergoes an AFM$_\text{S}$-PM$_\text{S}$-PM$_\text{L}$ type quantum phase transition, where the magnetic order quenches prior to Kondo-breakdown transition, and the Fermi surface reconstructs within the paramagnetic state.
    }
\end{flushleft}
\normalsize
\end{figure*}

\end{document}